\documentclass[journal]{IEEEtran}
\usepackage{makecell}
\usepackage{hyperref}
\usepackage{array}
\usepackage{graphicx,amssymb,amsmath}
\usepackage{multicol}
\usepackage[noadjust]{cite}
\usepackage{setspace}
\usepackage{subfigure}
\usepackage{graphicx}
\usepackage{float}
\usepackage {url}
\usepackage{stfloats}
\usepackage{amsthm,pifont}
\usepackage{flushend}
\usepackage{cases,subeqnarray}
\usepackage{bm,multirow,bigstrut}
\usepackage{amsmath, amsthm, amssymb}
\usepackage{textcomp}
\usepackage{latexsym,bm}
\usepackage{booktabs}
\usepackage{xcolor}
\usepackage{mathtools}
\usepackage{dsfont}
\usepackage{extarrows}
\usepackage{epsfig}
\usepackage{epstopdf}
\usepackage{algpseudocode}
\usepackage{algorithmicx,algorithm}
\usepackage{pdfpages}
\usepackage{tabularx}

	\usepackage{xspace}
	\usepackage{listings}
	\definecolor{commentcolor}{RGB}{85,139,78}
	\definecolor{stringcolor}{RGB}{206,145,108}
	\definecolor{keywordcolor}{RGB}{0,0,128}
	\definecolor{backcolor}{RGB}{220,220,220}

\begin{document}

\title{YOLO-based Semantic Communication with Generative AI-aided Resource Allocation for Digital Twins Construction}

\author{{Baoxia Du, Hongyang Du, Haifeng Liu, Dusit Niyato,~\IEEEmembership{Fellow,~IEEE,} Peng Xin, Jun Yu, Mingyang Qi, You Tang}

\thanks{This work was supported by Jilin Scientific and Technological Development Program ({YDZJ202201ZYTS692}).}
\thanks{B. Du, Y. Tang are with School of Electrical and Information Engineering, JiLin Agricultural Science and Technology University, Jilin 132101, China, and also with School of Information and Control Engineering, Jilin Institute of Chemical Technology, Jilin 132022 (e-mail: dubaoxia@jlict.edu.cn and tangyou9000@163.com).}
\thanks{H. Du, and D. Niyato are with School of Computer Science and
Engineering, Nanyang Technological University, Singapore (e-mail: hongyang001@e.ntu.edu.sg and dniyato@ntu.edu.sg).}
\thanks{H. Liu is with College of Agriculture, Yanbian University, Yanji 133002 (e-mail: liufeng\_1989@163.com).}
\thanks{P. Xin and J. Yu are with School of Information and Control Engineering, Jilin Institute of Chemical Technology, Jilin 132022 (e-mail: xinpeng4321@163.com and yujun@jlict.edu.cn).}
\thanks{M. Qi is with School of Electrical and Information Engineering, JiLin Agricultural Science and Technology University, Jilin 132101, China (e-mail: qimingyang0912@126.com).}
\thanks{Corresponding author: M. Qi and Y. Tang.}
}




\maketitle

\begin{abstract}
Digital Twins play a crucial role in bridging the physical and virtual worlds. Given the dynamic and evolving characteristics of the physical world, a huge volume of data transmission and exchange is necessary to attain synchronized updates in the virtual world. In this paper, we propose a semantic communication framework based on You Only Look Once (YOLO) to construct a virtual apple orchard with the aim of mitigating the costs associated with data transmission. Specifically, we first employ the YOLOv7-X object detector to extract semantic information from captured images of edge devices, thereby reducing the volume of transmitted data and saving transmission costs. Afterwards, we quantify the importance of each semantic information by the confidence generated through the object detector. Based on this, we propose two resource allocation schemes, i.e., the confidence-based scheme and the artificial intelligence-generated scheme, aimed at enhancing the transmission quality of important semantic information. The proposed diffusion model generates an optimal allocation scheme that outperforms both the average allocation scheme and the confidence-based allocation scheme. Moreover, to obtain semantic information more effectively, we enhance the detection capability of the YOLOv7-X object detector by introducing new  Efficient Layer Aggregation Network-HorNet (ELAN-H) and SimAM attention modules, while reducing the model parameters and computational complexity, making it easier to run on edge devices with limited performance. The numerical results indicate that our proposed semantic communication framework and resource allocation schemes significantly reduce transmission costs while enhancing the transmission quality of important information in communication services. 
\end{abstract}

\begin{IEEEkeywords}
Semantic communication, resource allocation, object detection, digital twins.
\end{IEEEkeywords}

\section{Introduction}

In recent years, the advances in technologies such as Augmented/Extended/Virtual Reality (AR/XR/VR), blockchain, Sixth-Generation (6G) network, artificial intelligence (AI) and edge computing have led to an increasing demand for virtual reality and digital worlds. The metaverse, as a virtual reality concept, is considered an integration of multiple virtual worlds that can provide people with a more immersive and realistic digital space~\cite{27Chu2022MetaSlicingAN,10098667VirtualReality}, allowing them to engage in various activities such as virtual conferences, remote collaboration, online learning, digital exhibitions, etc. The emergence of these virtual reality activities has not only alleviated social isolation and transportation restrictions, but also saved time and costs, gradually becoming essential tools for people's lives and work. Moreover, the popularity of these activities in various social domains has also accelerated the development of digital economy and digital transformation~\cite{6Wu2021DigitalTN,2Feng2022theMF}.

In agriculture, novel paradigms such as digital farms, smart agriculture, and agricultural metaverse, which are combined with metaverse technology, are emerging and flourishing~\cite{2Feng2022theMF,Neethirajan2021DigitalTI}. In terms of agricultural production, users, such as farmers, can establish virtual farms in a virtual environment, simulate the complete growth process of crops and livestock, and obtain real-time growth data to achieve intelligent and refined agriculture. For instance, the XR Lab of Alibaba DAMO Academy presented a case study of an autonomous agricultural picking robot. The proposed approach entails the development of a high-precision Three-Dimensional (3-D) model of the entire orchard via 3-D modeling techniques of both the orchard and fruit trees. Subsequently, a motion planning scheme can be established in the virtual environment, which can facilitate the robot's autonomous picking process in the real world. This innovative approach can potentially minimize the costs associated with orchard management~\cite{ali}. Furthermore, virtual agriculture can be combined with other fields such as agricultural leisure and agricultural education. For example, the Faculty of Agriculture has developed an agricultural metaverse teaching system for an egg chicken farm at the National University of Laos. Through VR technology, the faculty members provide agricultural education to learners, including knowledge related to technology, farm location, and other relevant aspects. The system has been reported to have yielded positive results~\cite{1Khansulivong2022AdaptiveON}.

Digital Twins (DTs), namely digital replications of physical objects, have emerged as a pivotal technology for creating virtual environments~\cite{4Han2023ADH,5Khan2021DigitalTwinEnabled6V}. In agriculture, the physical realm is characterized by its intricate and constantly-changing nature, necessitating DTs synchronizing with the physical world to ensure their accuracy in virtual operations. This process requires  that edge devices persistently gather the most recent data from the physical world, enabling real-time DTs updates. The acquisition and transmission of the data from the physical world often rely on various advanced edge devices and wireless communication technologies~\cite{7Farooq2019ASO,8Paraforos2021DigitalFA}. Various fixed or mobile devices (e.g., sensors and cameras) are deployed to collect status data of physical objects, which, in real-time, update and interact with the virtual world through wireless communication. However, continuous data transmission poses stringent requirements on wireless communication systems, especially when dealing with extensive data such as high-definition images, which can be both expensive and challenging when the physical world is vast.
	
Fortunately, semantic communication has been introduced as a novel avenue for tackling the aforementioned challenge~\cite{9Strinati20206GNB,10Zhang2021TowardWA}. In contrast to conventional communication technologies, semantic communication systems regard transmission effective if the meaning of the received information maintains the original meaning of the transmitted information~\cite{11Ng2022StochasticRA}. For example, in the context of image transmission, a semantic-based communication system can reduce the amount of data that needs to be transmitted by only transmitting the semantic information behind the image, while achieving the same effect~\cite{semdong}.
	
In this paper, we present a case study focusing on the development of a virtual apple orchard using a real apple dataset. In the virtual orchard, users, i.e., fruit growers, can easily access various information such as the quantity and location of fruit on each apple tree, as well as growth status and view real images of individual apples. The virtual orchard can help users manage their orchard more efficiently. In this case, the implementation of DTs requires edge devices, such as Unmanned Aerial Vehicle (UAV) to capture the status information of fruit trees by taking photographs, and then transmitting the collected data to users via wireless communication technology. To reduce costs and enhance communication quality during this process, we propose a semantic communication and resource allocation framework based on You Only Look Once (YOLO). Our main contributions are summarized as follows:
\begin{itemize}
\item {Unlike traditional communication methods that necessitate transmitting all acquired images, we propose a YOLO-based semantic communication framework. Specifically, the proposed framework discards irrelevant interference information after image data acquisition, retaining only the critical semantic information for transmission. This significantly reduces the data volume needed for transmission and lowers resource costs while achieving the same outcome.}
\item {We employ the YOLOv7-X object detector to extract semantic information from images and enhance its performance on a real-world apple dataset. Considering the limitations of existing object detectors in detecting small objects such as small apples, and the constraints of processing power and memory in edge devices, larger models necessitate increased computational resources and memory for operation, which may result in performance degradation or inoperability. Consequently, we improve the YOLOv7-X algorithm by introducing the Efficient Layer Aggregation Network-HorNet (ELAN-H) and the SimAM attention modules. These modifications elevate the detector's performance and reduce the parameters and computational requirements, facilitating deployment on edge devices with greater ease.}
\item {In the pursuit of enhancing transmission quality during the wireless transmission of a huge volume of images, we propose a resource allocation scheme based on the significance of semantic information. The scheme allocates transmission power following the relative importance of the semantic information, with the aim of enhancing the overall communication quality of image transmission systems by minimizing important information loss and improving the reliability of transmitted information. This approach ensures that critical information is transmitted with high quality to users, even in challenging wireless communication environments.}
\item {Furthermore, we utilize the AI-generated resource allocation scheme algorithm as an alternative allocation scheme, which facilitates more efficient processing power distribution. Specifically, by using the denoising technique, the AI-generated algorithm generates a design for the allocation scheme and subsequently adds exploration noise to execute it, thereby gaining experience in exploration. The numerical results clearly demonstrate that this method achieves the highest score in terms of semantic information transmission quality.}
\end{itemize}

The remainder of this paper is organized as follows: In Section~\ref{S2}, we initially summarize the related work about DTs, semantic communication and apple detection. Section~\ref{S3} introduces the overall system design, semantic communication approach, and the metric used for evaluating the system's communication quality. In Section~\ref{S4}, the YOLOv7-X object detector and its enhancement methods are described in detail, followed by an explanation of two distinct resource allocation methods for data transmission. Subsequently, we analyze the numerical results in Section~\ref{S5}. Lastly, Section~\ref{S6} concludes the paper.
	\section{Related Work}\label{S2}
    In this section, we briefly introduce three related techniques, i.e., DTs, semantic communication and apple detection.
        \subsection{Digital Twins}\label{S21}
     The physical system and physical world in agriculture are complex and dynamic environments that include basic information and characteristics of physical objects. DTs require continuous updating from the physical to the virtual space as the state of physical objects changes over time~\cite{11Han2022ADH,10002946iot}. Li et al.~\cite{12Li2022ASV} proposed a deep learning-based single-view leaf reconstruction method for a plant growth DT system, improving leaf reconstruction's accuracy and speed. Angin et al.~\cite{13Angin2020AgriLoRaAD} introduced a DT framework for agriculture called AgriLoRa, which detects plant diseases and weeds using computer vision algorithms after uploading data from UAV images and field sensor data to cloud servers. Awais et al.~\cite{14Awais2022OptimizationOI} used the multispectral UAV and DTs model to achieve intelligent irrigation in the field. However, these works have focused only on the use of collected data and have not considered the impact of data transmission. DTs require a significant amount of computing power to render 3-D objects, which is achieved through collecting large amounts of data from perception networks and ultra-low latency communication to maintain a seamless user experience~\cite{27Chu2022MetaSlicingAN}. When physical objects are large enough, the massive data streams can burden communication systems and cause excessive latency, or even transmission failure. Therefore, in this paper, we use a semantic-aware communication method to reduce the amount of data that needs to be transmitted.
        \subsection{Semantic Communication}\label{S22}
    In classical communication theory, the semantic content and meaning of the message are largely considered irrelevant to communication. However, in the age of rapidly increasing data volume, the limitations of classical communication theory have begun to be revealed~\cite{28Lan2021WhatIS}. Semantic communication constitutes an innovative paradigm wherein message transmission is not confined solely to the message content, but rather entails direct extraction of pertinent semantic information, thereby eliminating redundant data and mitigating associated costs. Xie et al.~\cite{29Xie2020DeepLE} proposed a text transfer framework called DeepSC based on the Transformer~\cite{36Vaswani2017AttentionIA}, which can recover the meaning of sentences through semantic information, thus minimizing semantic errors during transmission. Zhou et al.~\cite{30Zhou2022CognitiveSC} proposed a cognitive semantic communication framework that utilizes knowledge graphs, which has good data compression rates and communication reliability. In addition to text-based semantic communication, some literature also proposes semantic communication methods applied to images.~\cite{31Lokumarambage2023WirelessEI} obtained semantic information through semantic segmentation at the transmitter and used GAN networks to reconstruct the image at the receiver, greatly saving bandwidth resources, but the reconstructed image is slightly different from reality. Zhang et al.~\cite{32Zhang2022DeepLS} proposed a neural network-based image transfer semantic communication system, where the transmitter can extract and transmit the required semantic information in a dynamic environment through a receiver-leading training process without knowing the task. Kang et al.~\cite{33Kang2022PersonalizedSI} proposed a task-oriented semantic communication framework, where users can match the semantic information of images by querying text, and also consider the resource allocation problem when there are multiple users.
    Although the above literature reduce communication overhead through semantic communication, they have not considered the varying importance of semantic information itself, which may result in the loss of significant semantic information in the competition for channel resources. Therefore, we assign different levels of importance to the semantic information extracted from images to ensure the transmission quality of critical semantic information by rationally allocating transmission power.
 	\subsection{Object Detection}\label{S23}
    In the communication framework proposed in this paper, the UAV needs to extract the semantic information of the acquired images, i.e., to achieve the separation of apples and backgrounds, and the core of achieving this is the object detection technique. 
    
    \textbf{Apple Detection: }In recent years, deep learning-based object detection techniques have achieved remarkable success. In contrast to traditional algorithms that rely on appearance features such as shape and color~\cite{12Lv2016RecognitionOO,13Liu2019ADM}, deep learning-based techniques demonstrate strong adaptability to different scenarios and achieve higher accuracy. Chen et al.~\cite{20Chen2021AnAD} utilized the DenseNet network structure to optimize the YOLOv4 model, proposing a Des-YOLOv4 algorithm for detecting apples. However, the performance of the algorithm significantly deteriorates under low-light conditions. Yan et al.~\cite{21Yan2021ARA} proposed an improved YOLOv5~\cite{yolov5} algorithm for real-time apple recognition by incorporating Squeeze-and-Excitation (SE) modules and modifying the fusion mode of feature maps. Despite the improved performance, the algorithm's effectiveness in detecting small apples is suboptimal. Sun et al.~\cite{22Sun2022BFPNB} proposed a novel Balanced Feature Pyramid Network (BFP Net) that enhances the accuracy of small apple detection. Nevertheless, the BFP Net has a slower detection speed. The above work indicates that it is difficult to balance the detection speed and accuracy of the model, and detecting small apples in complex environments remains a challenge. Therefore, we select the YOLOv7-X object detector for this study due to its outstanding detection speed and accuracy~\cite{23Wang2022YOLOv7TB}. However, its performance on dataset with a considerable number of small apples is slightly limited. Thus, we employed it as the baseline model for improving detection performance.

    \textbf{Data Augmentation: }In addition to modifying the model structure to improve detection capability, data augmentation is another straightforward and effective approach. Data augmentation allows generating additional equally effective data based on limited data without altering the essential information of the images. This significantly enhances the diversity of the training data, thereby enhancing the model with stronger generalization capabilities. Changing the color and shape of images is the fundamental and common approach in data augmentation. In this study, we use fundamental data augmentation methods such as randomly altering the hue, saturation, and brightness of images, as well as performing random scaling and translation. In addition to these basic data augmentation methods, some studies have proposed more efficient approaches, such as Mosaic~\cite{24Bochkovskiy2020YOLOv4OS} and Mixup~\cite{25Zhang2017mixupBE}. Mosaic involves randomly cropping and scaling four images, then combining them into a single image for training data. Mixup randomly selects two samples from the training data and constructs new training samples and labels through linear interpolation. In this work, we use a combination of Mosaic, Mixup and fundamental methods.

    In summary, we propose a YOLO-based semantic communication and resource allocation framework to address the above problems in this paper. First, we select the advanced YOLOv7-X object detector as the core of the entire system and optimize its performance on real dataset. Following that, the UAV extracts semantic information from collected images using the optimized YOLOv7-X object detector. Then, the UAV allocates transmission power based on the importance of semantic information before transmitting, ultimately achieving high-quality transmission of critical information.
    
        \section{System Model}\label{S3}
    In this section, we present the proposed data transmission framework and the semantic-based communication approach, followed by an explanation of the metric methodology for assessing image semantic information transmission quality.
        \subsection{Design of  semantic communication and resource allocation framework}\label{S31}
\begin{figure*}[ht]
	\centering
	\includegraphics[width=0.9\textwidth]{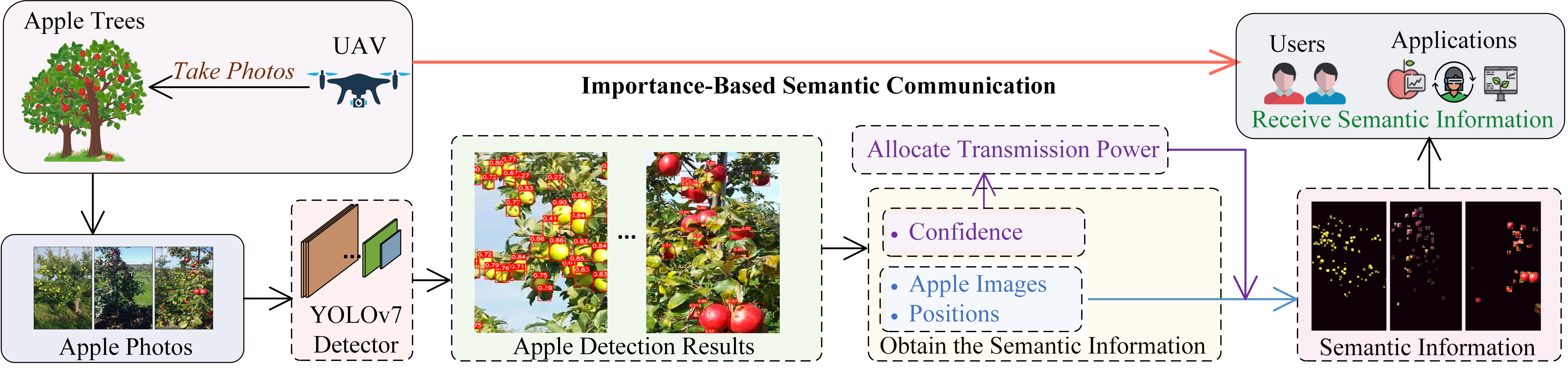}%
	\caption{An illustration of the YOLO-based semantic communication system model.}
	\label{frame}
	\end{figure*}

    In data collection in DTs, a UAV takes images along a specified trajectory and transmit the data to users. However, in wireless communication environments with fading channels, the cost of transmitting all data to users at high quality is prohibitive. To address this issue, we propose a semantic communication framework based on YOLO, as shown in Fig. \ref{frame}. We first simulate the images captured by the UAV using real apple dataset. After the collection task is completed, the UAV uses the trained YOLOv7-X object detector to extract semantic information behind the image that people need and transmit them. At the same time, we quantify the importance of each semantic information based on the confidence generated by the object detector and allocate transmission power accordingly to ensure the transmission quality of important information. The problem of optimal power allocation schemes is introduced in Section~\ref{S4}. It is worth noting that the YOLOv7-X object detector is a flexible module that can be applied to semantic communication systems in various scenarios by training on different datasets. For instance, use pretrained models to extract semantic information from images containing other single or multiple different classes of objects.
        \subsection{Semantic Communication Solution}\label{S32}
    In conventional communication paradigms, edge devices transmit the entirety of the acquired image data to facilitate continuous data synchronization for DTs, resulting in voluminous data traffic. This imposes substantial burdens on both edge devices and the communication infrastructure. To illustrate, in smart agriculture, users deploy UAVs for image capture~\cite{9903855smart}. However, oftentimes, only a portion of the captured content is relevant to users, e.g., ripening fruits. Utilizing such an inefficient communication approach for transmitting all images leads to the excessive consumption of communication resources and energy for UAVs, thereby exacerbating resource wastage.
    
	The development of semantic communication provides a solution to the aforementioned problems. Upon completing the designated data acquisition task, the UAV transmits only the pertinent semantic features extracted from the captured images, rather than the entire dataset. The transmission of these semantic features requires minimal channel resources and facilitates efficient data storage for users. In the virtual orchard, users, e.g., fruit growers, are primarily more concerned about the status of the fruits. As a result, we discard irrelevant background and interference factors before transmitting the images, ensuring that users only receive the semantic information they are interested in, which helps reduce transmission cost.
 
        \subsection{Semantic Communication Quality Analysis}
    To evaluate the quality of semantic communication, we propose a Metric for Image Semantic Transmission (MIST) in this work, which involves combining the importance weights of each semantic information with their respective transmission quality to obtain the final evaluation result. Considering that a UAV needs to send an image to the user after capturing it, semantic information is first extracted by the object detector. Specifically, a total of $U$ apple objects are detected, with $i$ denoting the $i_{\rm th}$ object and ${c_i}$ denoting its corresponding confidence. The relationship between the importance score ${W_i}$ and the confidence  ${c_i}$ of the object $i$ can be represented as ${W_i} = {c_i}^{\sigma},$ where $\sigma $ is a variable that adjusts the importances among different semantic information. The final semantic transmission quality score can be represented as follows:      
         \begin{equation}\label{reward}
         E(A,{W_i},Q({p_{\rm{i}}})) = A\sum\limits_{i = 1}^U {({W_i} \times Q({p_{\rm{i}}}))} ,
         \end{equation}
    where $A$ represents the accuracy of semantic information extraction, i.e., the performance evaluation metric Average Precision at 0.5 Intersection over Union (AP@0.5) of the object detector, and $Q({p_{\rm{i}}})$ represents the Structure Similarity Index Measure (SSIM)~\cite{47Wang2004ImageQA} value of object $i$ before and after transmission, which is a function that is positively correlated with the transmission power ${p_i}$ according to~\cite{33Kang2022PersonalizedSI}. Therefore, considering  the definition of the communication system in this paper, our goal is to maximize the MIST while satisfying the transmission power constraint, which can be expressed as follows:

        \begin{align}  
        \mathop {\max }\limits_{A,{W_i},{p_i}} \sum\limits_{i = 1}^U &{E(A,{W_i},Q({p_{\rm{i}}})))}, \label{YY}\\ 
        \sum\limits_{{\rm{i}} = 1}^U &{{p_i}}  \le P,      
        \tag{\ref{YY}{a}} \label{YYa}\\
        {{\rm{c}}_i} \in &[{{\rm{c}}_{\min }},1], \tag{\ref{YY}{b}} \label{YYb}
        \end{align}        
    where the constraint in \eqref{YYa} is the total transmitted power, and ${c_{\min }}$ in \eqref{YYb} is the confidence threshold, the objects with confidence below this value are not detected by the detector. In this study, we set the variable ${\sigma } = 1$ and the confidence threshold ${c_{\min }} = 0.25$ by default. The proposed MIST considers not only the transmission quality of each semantic information but also their respective significance, which can provide a more comprehensive and accurate assessment of the performance of these methods.
    
        \section{YOLO-based Semantic Communication System Design}\label{S4}
    The YOLOv7-X object detector is a critical component of the overall communication system. It aids UAVs in extracting semantic information from images and subsequently quantifying the semantic information's importance to facilitate optimal transmission power allocation. In this section, we provide a detailed exposition of the YOLOv7-X, along with the improvements made to it.  Subsequently, we discuss two distinct resource allocation schemes in data transmission.
        \subsection{Overview of YOLOv7-X}\label{S41}
    The YOLOv7 model is a highly advanced and efficient end-to-end object detector. It employs a state-of-the-art methodology for detecting objects in an image, with exceptional accuracy and real-time performance. YOLOv7 has seven different models of varying sizes, including YOLOv7-tiny, YOLOv7, YOLOv7-X, and YOLOv7-W6, among others, which are suitable for different application environments. Considering both model complexity and detection performance, we select YOLOv7-X as the base model for improvement. The YOLOv7-X model can be divided into three parts: Input, Backbone, and Head. Specifically, the Input resizes the input image to the required training size. The Backbone component includes multiple CBS convolutions, Max Pooling Convolution (MPConv), Efficient Layer Aggregation Network in YOLOv7-X (ELAN-X) modules and an SPPCSPC module. The CBS convolution consists of a convolutional layer, a Batch Normalization layer (BN), and a Sigmoid Linear Unit (SiLU) activation function. ELAN-X extends the ELAN module by increasing its depth and width, and enhances the learning capability of the network by guiding the computation blocks to learn more diverse features of different feature groups. The MPConv module adds a Maxpool layer on top of CBS and strengthens the feature extraction ability by merging features from the top and bottom branches. The SPPCSPC module is similar to the Spatial Pyramid Pooling-Fast (SPPF) used by YOLOv5~\cite{yolov5}, which increases  a network's receptive field. The Head component employs the same Path Aggregation Feature Pyramid Network (PAFPN)~\cite{34Ge2021YOLOXEY} architecture as YOLOv5 to efficiently fuse features from multiple levels. Finally, the fused and enhanced feature map is fed to three detection heads to generate predictions for confidence, object category, and anchor boxes.
        \subsection{Model Enhancement Methods}\label{S42}
\begin{figure}[ht]
        \centering 
        \subfigure[ELAN-H model structure, which stacks feature maps from different levels by Concatenation (Concat) operation as the final output.]{
        \label{elanh}
        \includegraphics[width=0.25\textwidth]{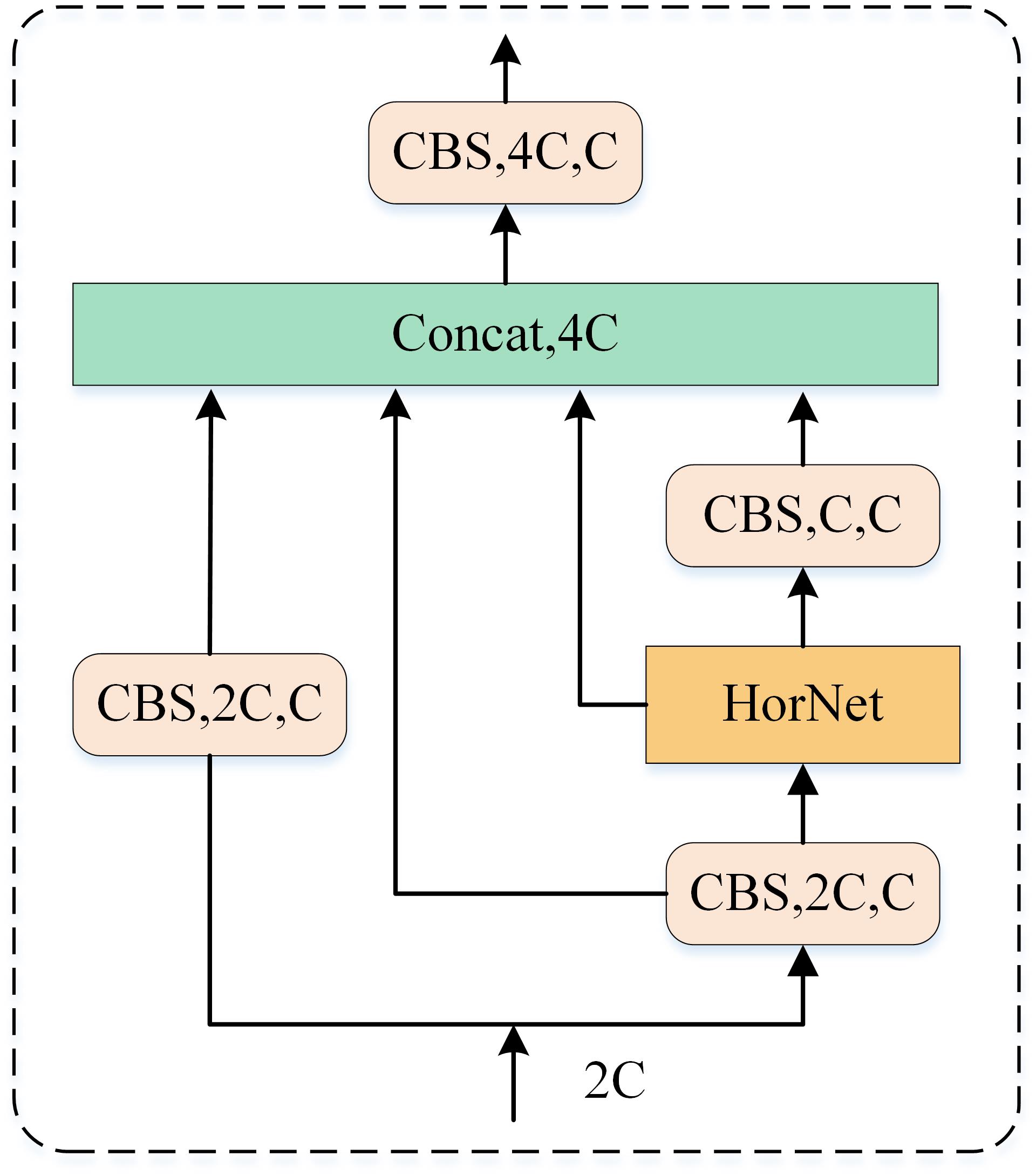}}
        \subfigure[Overview of the basic building block in HorNet with Recursive Gated Convolution (${g^n}{\rm{Conv}}$).]{
        \label{gnconv}
        \includegraphics[width=0.3\textwidth]{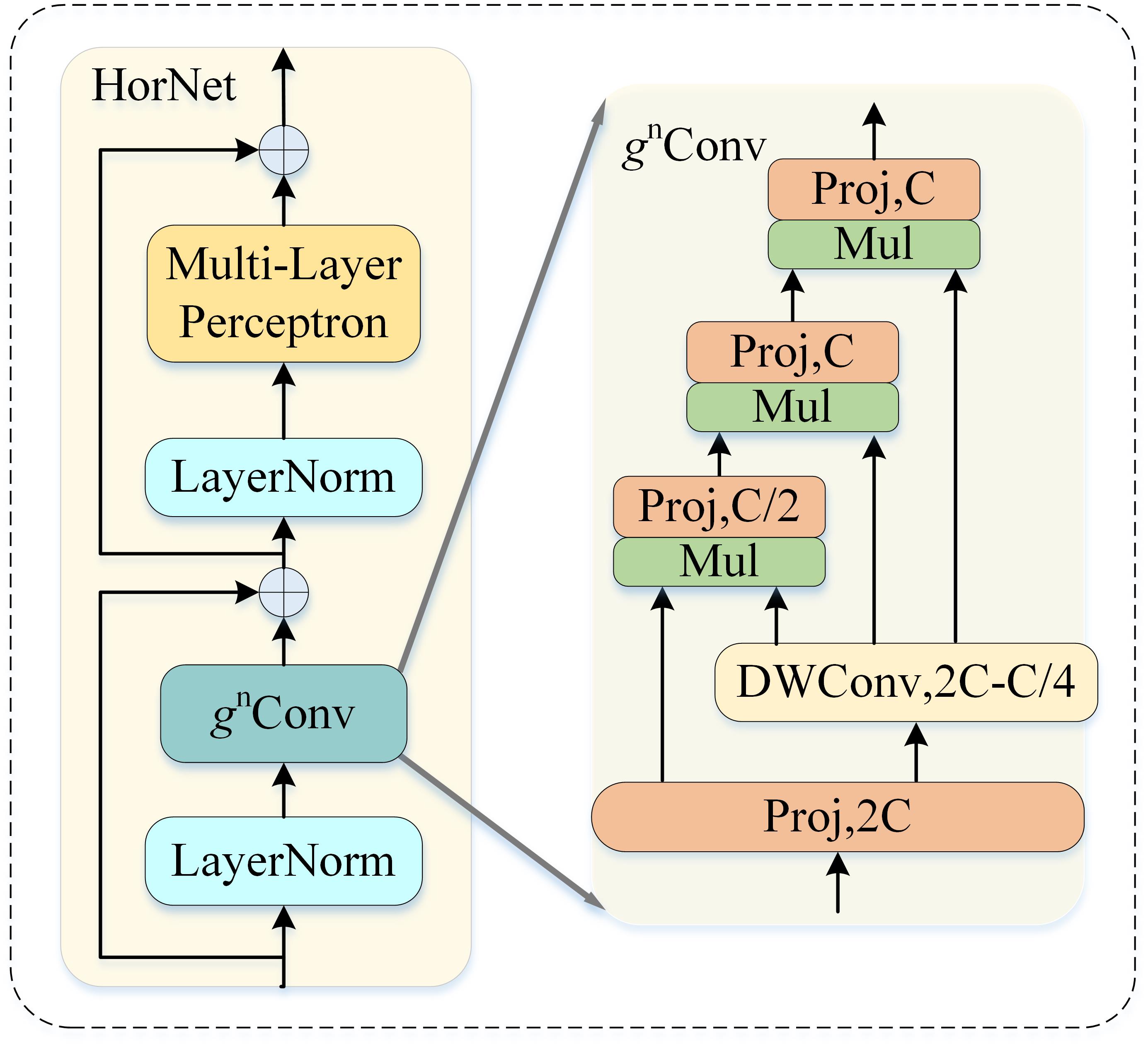}} 
        \caption{ELAN-H model structure with HorNet.}
        \label{enh}
        \end{figure}

\begin{figure*}[htbp]
	\centering
	\includegraphics[width=0.95\textwidth]{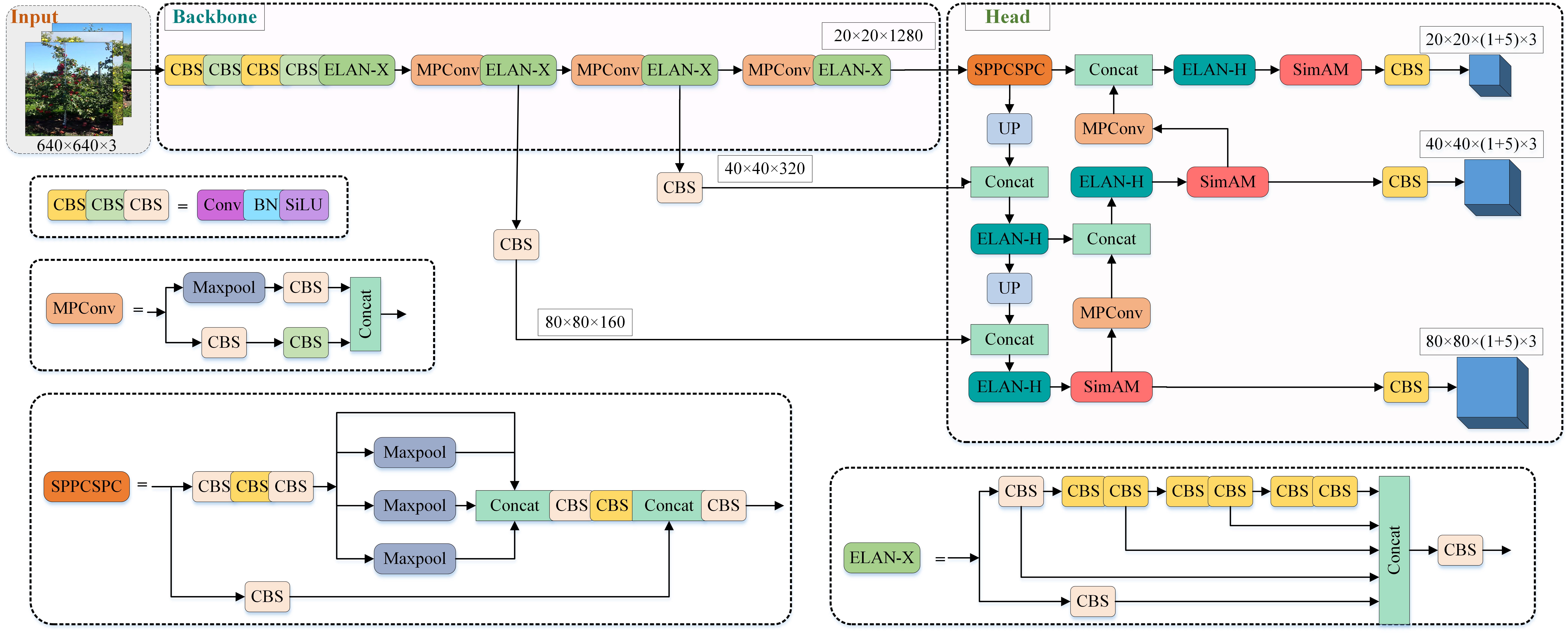}
	\caption{The network architecture of YOLOv7-HS contains general modules: Input, Backbone, and Head, and basic components: CBS, Max Pooling Convolution (MPConv), UPSampling (UP), SPPCSPC, SimAM, and two Efficient Layer Aggregation Networks (ELAN) with different structures, namely ELAN-X and ELAN-H.}
	\label{yolov7}
	\end{figure*}

    \textbf{ELAN-H: }The ELAN-X module in YOLOv7-X is an efficient network structure that enables the network to learn more features and have stronger robustness by controlling the shortest and longest gradient paths. The structure of ELAN-X is shown in Fig. \ref{yolov7}. The shortest branch passes through only one CBS convolution to change the number of channels, while the longest branch extracts features through seven CBS convolutions. The feature maps extracted by each branch are concatenated through the Concatenation (Concat) operation as the final result of feature-enhanced fusion. ELAN-X is repeatedly used in the Neck to improve the model's learning capability. However, too many branches and convolution operations also increase model complexity and parameter size, leading to increased processing time and consumption of computing resources. Therefore, we reduce the depth and width of this module without breaking its original architecture, changing the number of CBS in the longest branch to three and correspondingly decreasing the number of output feature maps to four. In addition, to compensate for the decrease in detection performance caused by simplifying this module, we replace one CBS convolution in the long branch with Hornet Block \cite{HorNetEH} to enhance the module's ability to learn important features. Finally, this module, named ELAN-H, has a structure shown in Fig. \ref{elanh}.
        
    Here, the Hornet block is a design based on Recursive Gated Convolution (${{g}^{n}}\text{Conv}$). The output of the gated convolution ${\bf{y}} = g{\rm{Conv(}}{\bf{x}}{\rm{)}}$ can be represented as follows:
         \begin{equation}
         \left[ {{{\bf{p}}_0}^{HW \times C},{\bf{q}}_0^{HW \times C}} \right] = {\phi _{in}}({\bf{x}}) \in {{\rm{R}}^{HW \times 2C}},
         \end{equation}
         \begin{equation}
         {{\bf{p}}_1} = f({{\bf{q}}_0}) \odot {{\bf{p}}_0} \in {R^{HW \times C}},y = {\phi _{out}}({{\bf{p}}_1}) \in {R^{HW \times C}},
         \end{equation}
    where ${\bf{x}} \in {{\rm{R}}^{HW \times C}}$ represents the input features which are linearly projected to obtain ${{\bf{p}}_0}$ and ${{\bf{q}}_0}$. Then, ${{\bf{q}}_0}$ is subjected to depth-wise convolution and multiplied with ${{\bf{p}}_0}$ to obtain ${{\bf{p}}_1}$. Lastly, ${{\bf{p}}_1}$ is projected linearly again to yield the output ${\bf{y}}$.

    High-order spatial interaction requires the implementation of gated convolutions with recursive designs. Initially, a higher-order linear projection is applied to ${\bf{x}}$ resulting in ${{\bf{p}}_0}$ and ${{\bf{q}}_k}(k = 0,1,...,n - 1)$. Subsequently, recursive gated convolutions are executed to generate ${{\bf{p}}_{k + 1}}$. The output of ${\bf{y}} = {g^n}{\rm{Conv(}}{\bf{x}}{\rm{)}}$ can be mathematically expressed as follows:
         \begin{equation}
         \begin{split}       
         \left[ {{{\bf{p}}_0}^{HW \times {C_0}},{\bf{q}}_0^{HW \times {C_0}},...,{\bf{q}}_{n - 1}^{HW \times {C_{n - 1}}}} \right] \\ = {\phi _{in}}({\bf{x}}) \in {{\rm{R}}^{HW \times ({C_0} + \sum\nolimits_{0 \le k \le n - 1} {{C_k}} )}},
         \end{split}
         \end{equation}
         \begin{equation}
         {{\bf{p}}_{k + 1}} = {f_k}({{\bf{q}}_k}) \odot {g_k}({{\bf{p}}_k})/\alpha , \qquad  k = 0,1,...,n - 1,
         \end{equation}
         \begin{equation}
         {g_k} = \left\{ \begin{array}{l}
         {\rm{Identity,}}    \:\qquad\qquad\qquad k{\rm{ = 0,}}\\
         {\rm{Linear}}({C_{k - 1}},{C_k}), \quad\quad 1 \le k \le n - 1.
        \end{array} \right.,
         \end{equation}
         \begin{equation}
         {C_k} = \frac{C}{{{2^{n - k - 1}}}},  \quad\qquad\qquad 0 \le k \le n - 1,
         \end{equation}
    where $\{ {g_k}\} $ are utilized to match the dimension in various orders and $\{ {f_k}\}$ are depth-wise convolution layers. As depicted in Fig. \ref{gnconv}, the HorNet block employs a block-wise design inspired by Transformer~\cite{36Vaswani2017AttentionIA} and replace the self-attention sub-layer with ${{g}^{n}}\text{Conv}$ that have high-order spatial modeling capability. We replace all ELAN-X modules in the Neck with ELAN-H modules, which can better fuse and enhance the image features extracted by the Backbone, leading to improved detection performance while reducing model complexity.
    
\begin{figure}[htbp]
        \centering 
        \subfigure[Channel-wise attention.]{
        \label{sim1}
        \includegraphics[width=5.6cm,height = 2.8cm]{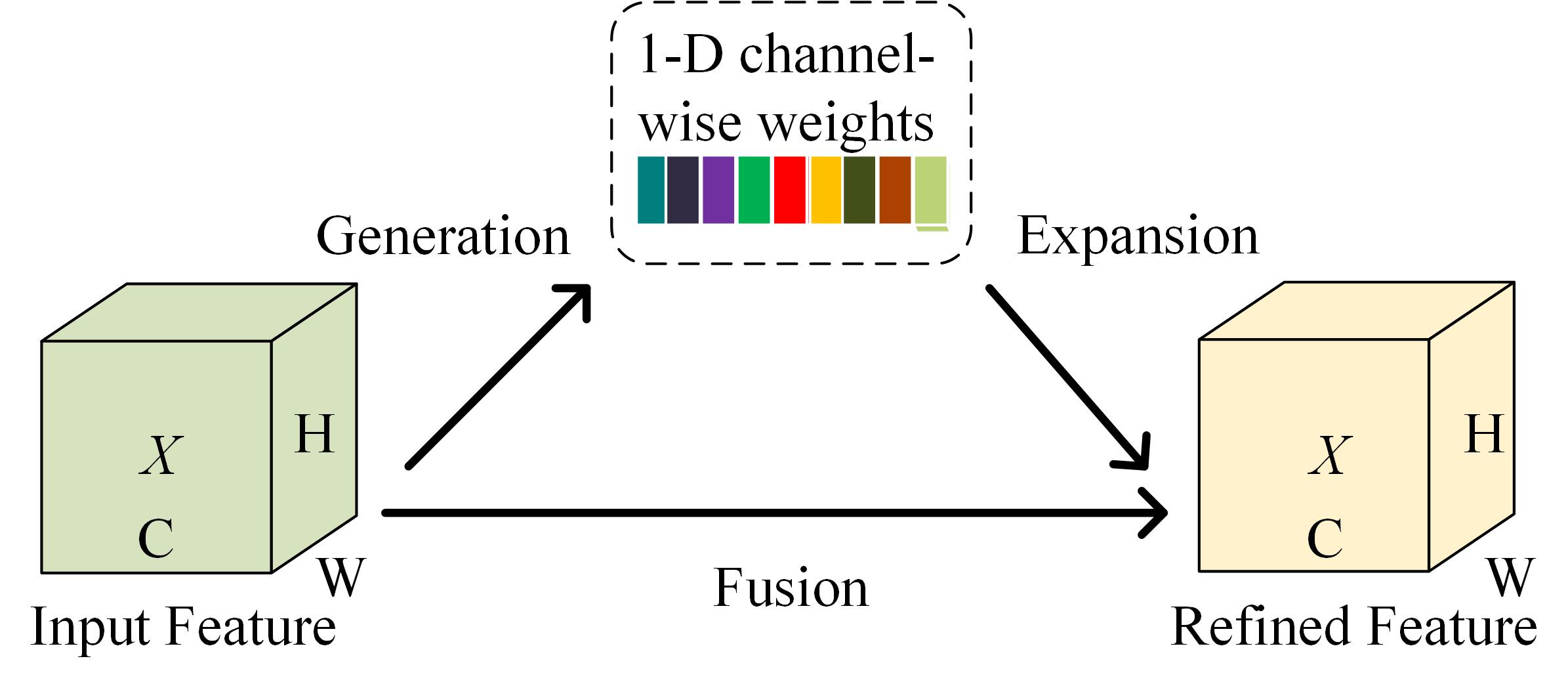}}
        \subfigure[Spatial-wise attention.]{
        \label{sim2}
        \includegraphics[width=5.6cm,height = 2.8cm]{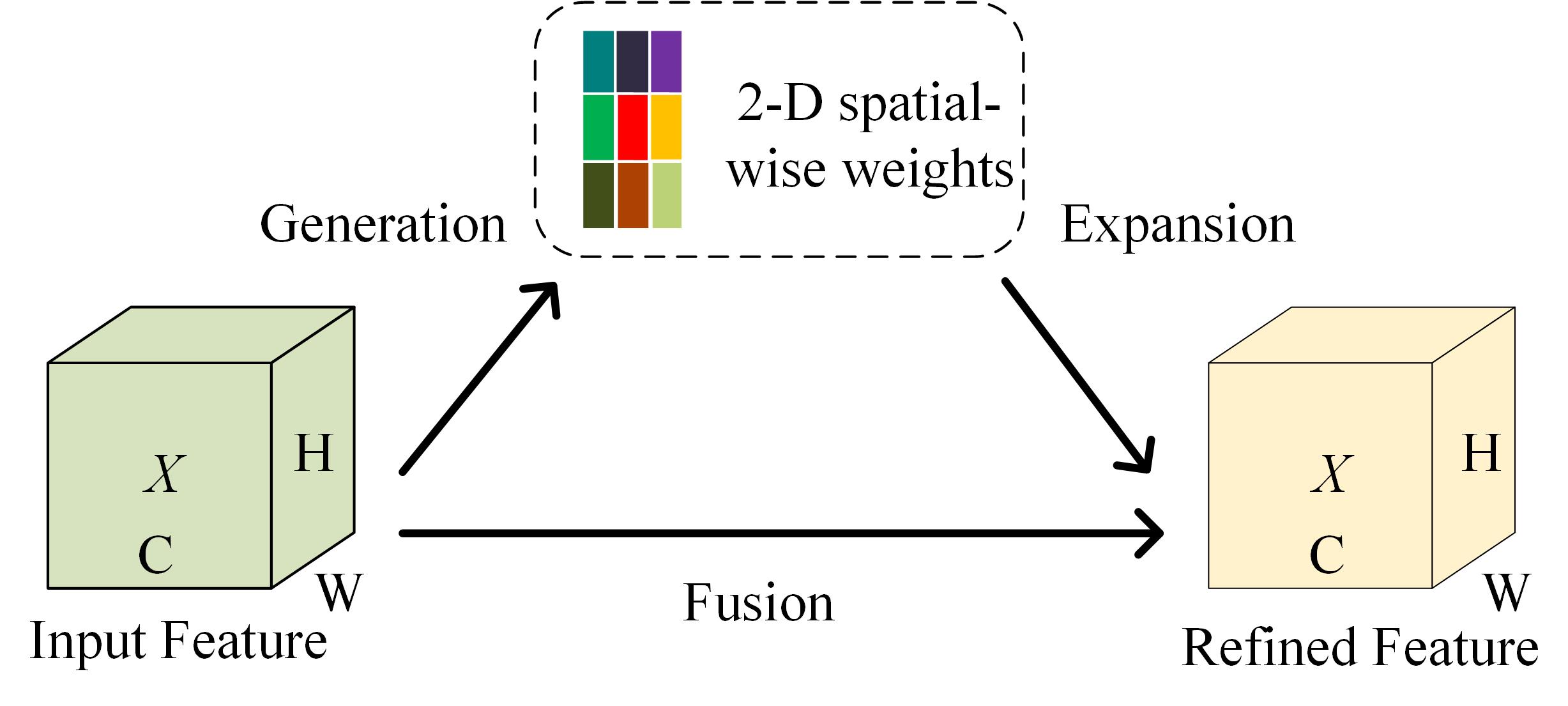}}
        \subfigure[Full 3-D weights for attention.]{
        \label{sim3}
        \includegraphics[width=5.6cm,height = 2.8cm]{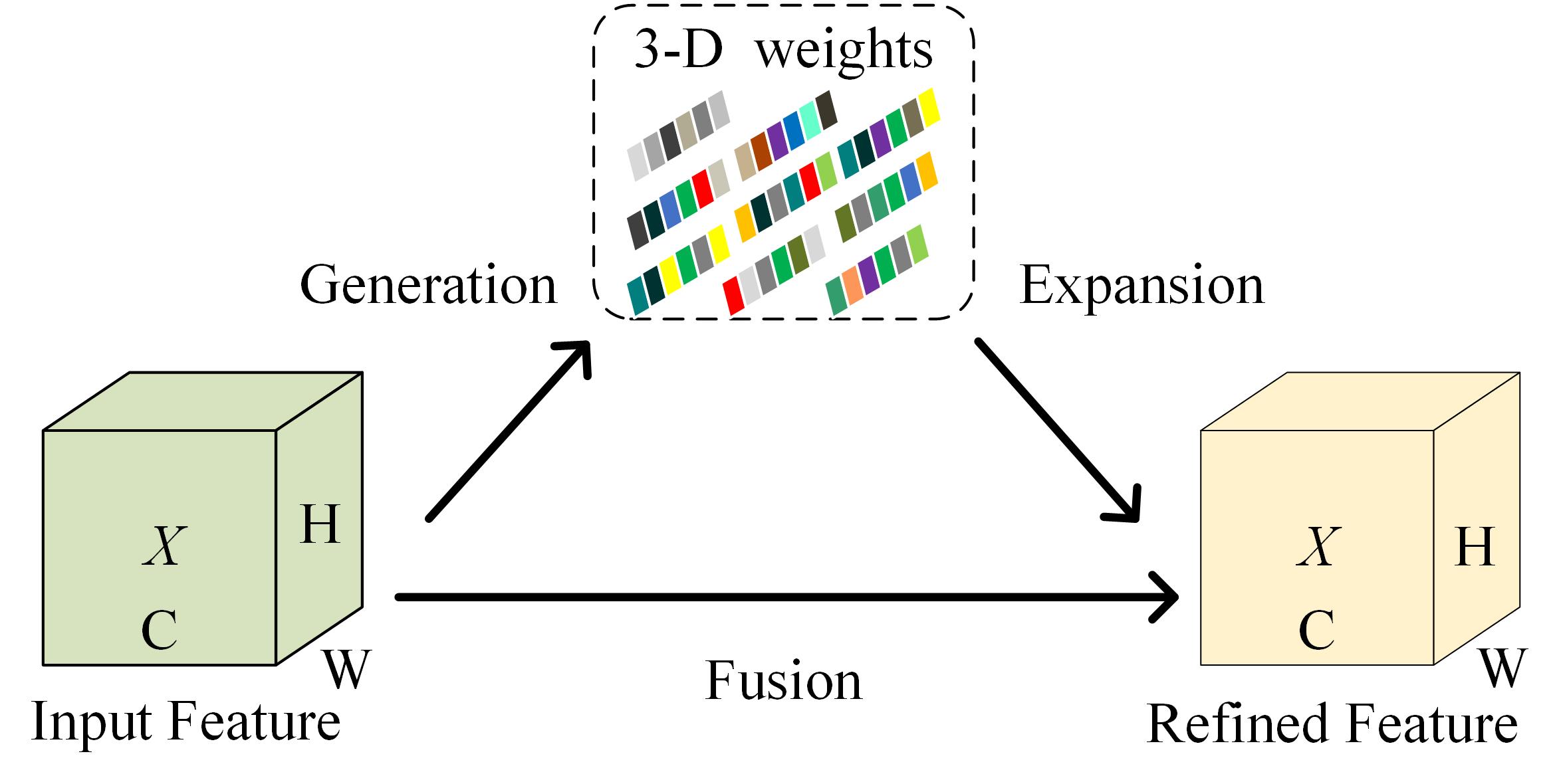}}
        \caption{Comparisons of different attention steps.}
        \label{sim}
        \end{figure}

    \textbf{SimAM~\cite{37Yang2021SimAMAS}: }The attention module can assign different weights to different channels or regions in space, thereby helping the model to focus on extracting more important information. Existing attention mechanisms typically generate corresponding 1-D or 2-D weights in the channel or spatial dimension, as shown in Fig. \ref{sim1} and Fig. \ref{sim2}, such as BAM~\cite{38Park2018BAMBA} , which parallelly connects two kinds of attention, and CBAM~\cite{39Woo2018CBAMCB}, which serially connects them. However, they treat each neuron in every channel or spatial position equally during the generation process. This limitation restricts their ability to learn more discriminative cues, while in the human brain, these two types of attention often occur simultaneously~\cite{37Yang2021SimAMAS}. 
    As shown in Fig.\ref{sim3}, SimAM is a unified weight attention module that can derive 3-D attention weights for feature maps without requiring additional parameters. In visual neuroscience, the most informative neurons typically exhibit discharge patterns that differ from those of surrounding neurons, and active neurons tend to inhibit surrounding neurons~\cite{40Webb2005EarlyAL}. Drawing inspiration from this, the SimAM module designs an energy function to measure the linear separability between neurons, thereby identifying important neurons. The energy function is defined as follows:
         \begin{equation}\label{eq1}
        \begin{split}
        e_t(w_t,b_t,y,x_i) &= \frac{1}{{M - 1}}\sum\limits_{i = 1}^{M - 1} {{{( - 1 - (w_tx_i + b_t))}^2}} \\
        &+ {(1 - (w_tt + b_t))^2} + \lambda w_t^2,
        \end{split}
         \end{equation}
    where $t$ denotes the target neuron and ${{x}_{i}}$ denotes other neurons in a single channel of the input feature $\mathbf{x}\in {{\text{R}}^{HW\times C}}$, $i$ is index over spatial dimension and $M$ is the number of neurons. $w_t$ and ${b_t}$ are weight and bias the transform. Subsequently, by computing the closed-form solutions of variables ${{w}_{t}}$ and ${{b}_{t}}$, and substituting them into \eqref{eq1}. The minimum energy can be obtained as follows:
         \begin{equation}
         e_t^* = \frac{{4({{\hat \sigma }^2} + \lambda )}}{{{{(t - \hat \mu )}^2} + 2{{\hat \mu }^2} + 2\lambda }}.
         \end{equation}         
    The above formula indicates that a smaller energy value corresponds to a greater linear separability between neuron $t$ and other neurons, which signifies higher importance. Following the definition of attention mechanisms, the SimAM module can be expressed as follows: 
         \begin{equation}\label{eq2}
         \tilde X = {\rm{sigmoid(}}\frac{1}{{\mathop{\rm E}\nolimits} }{\rm{)}} \odot {\rm{X}},
         \end{equation}
    where, $\operatorname{E}$ is a classification function for all energy functions $e_{t}^{*}$ in both channel and spatial dimensions. The sigmoid function ensures that the larger values in $\operatorname{E}$ are constrained, thereby not affecting the relative importance of each neuron. We integrate SimAM modules into the Neck section of the YOLOv7-X model, which helps the model to better focus on targets without introducing additional parameters.
    
    \textbf{YOLOv7-HS:} The improved architecture of the YOLOv7-X model is depicted  in Fig. \ref{yolov7}. We enhance the YOLOv7-X object detector using the above methods to make it more suitable for the Minneapple dataset, and named it YOLOv7-HS.
        \subsection{Resource Allocation in Semantic Communication}\label{S43}
    After the UAV obtains the semantic information of images, a straightforward approach is to transmit it equally to users. However, due to the limited bandwidth resources in wireless transmission, semantic information is susceptible to signal attenuation during transmission, consequently affecting communication quality. Simultaneously, as nearly every original image is cropped into numerous apple images, and considering the highly complex orchard environment and the inherent limitations of the detection model, the importance of each cropped image (e.g., apple completeness, cropping accuracy) varies. The average allocation method (named Avg-SemCom) faces intense competition for scarce wireless channel resources among multiple images, resulting in the discarding of crucial images. Therefore, we propose a Confidence-based Semantic Communication (Conf-SemCom) method, which allocates transmission power by quantifying the importance of semantic information to ensure the transmission quality of important data.

\begin{figure}[ht]
	\centering
	\includegraphics[width=0.46\textwidth]{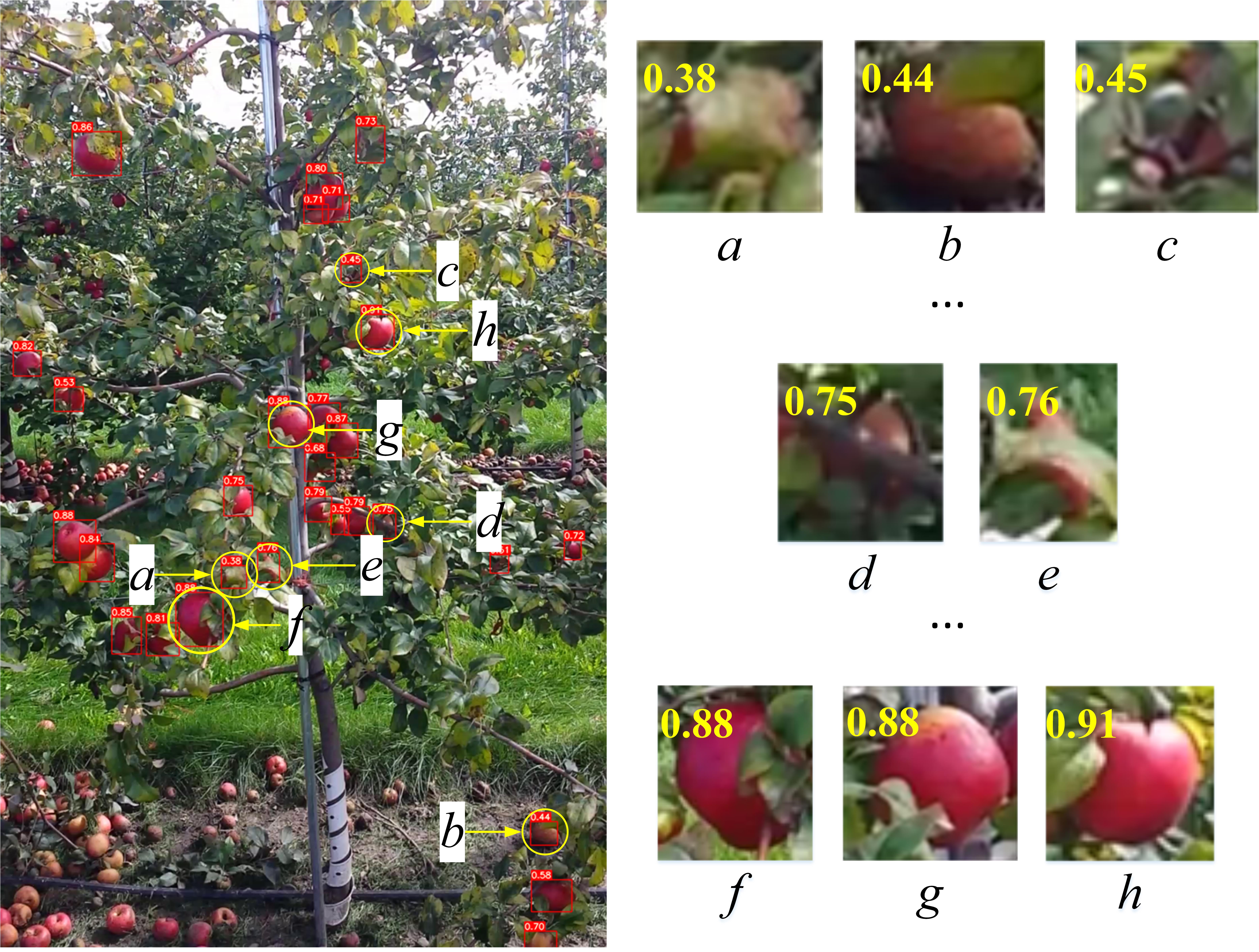}%
	\caption{The results of apple detection comprise anchor boxes and corresponding confidence values for each object, apples a to f are sorted in ascending order of confidence scores. Furthermore, the cropped images are resized to the same height for ease of viewing.}
	\label{apple351}
	\end{figure}

    The correlation between an object detector's confidence and the object it identifies is significant. Generally, the confidence of a detected object is positively correlated with the amount of semantic features it contains. As illustrated in Fig. \ref{apple351}, apples $f$, $g$, and $h$ exhibit relatively complete and rich characteristic information of apples, making them easily detectable with a high confidence. While apples $a$ and $c$ are heavily obscured by leaves, resulting in low confidence scores for all of them. The semantic information of these low-confidence objects, after being cropped, is far less important than the other objects, and they should not occupy too much power during wireless transmission. In summary, the Conf-SemCom method can ensure that more important information is successfully transmitted to users.

    Specifically, we prioritize sorting based on the confidence ${c_i}$ of each object ${i}$ and allocate more transmission power with a higher priority weight ${\bm{w}_i}$. The definition of ${\bm{w}_i}$ is as follows:
         \begin{equation}\label{wi}
         {\bm{w}_i} = {c_i}^\eta ,
         \end{equation}    
    we use the variable $\eta$ to adjust the relative difference in power allocation between different semantic information. The proposed Conf-SemCom is summarized in Algorithm \ref{Fills} with the corresponding pseudo code.
    
    	\begin{algorithm}[ht]
		\caption{Resource allocation in Conf-SemCom} 
		\label{Fills}
		\hspace*{0.02in} {\bf Input:}
		Captured image $X$ on a UAV\\
		\hspace*{0.02in} {\bf Output:}
		Users receive semantic information $x$ of the image $X$       
		\begin{algorithmic}[1]
                \Procedure{UAV-SEND($x$)}{}
   			\State YOLOv7-HS detects apple images ${x_1}$,${x_2}$,\ldots,${x_i}$,\ldots,${x_U}$ and their confidence ${c_1}$,${c_2}$,\ldots,${c_i}$,\ldots,${c_U}$
			\State Calculate the priority weight ${\bm{w}_i}$ of ${x_i}$ by confidence ${c_i}$ according to \eqref{wi}
			\For{$i = 1$ to $U$} 
			\State Allocate transmission power to ${x_i}$ according to their priority weight ${\bm{w}_i}$ using PA~\cite{33Kang2022PersonalizedSI}
			\EndFor
			\State The UAV sends semantic information $x$ to users
                \EndProcedure
		\end{algorithmic}
	  \end{algorithm}

    \subsection{Diffusion-Based Resource Allocation}\label{S44}
    \textbf{Diffusion model: }The diffusion model has emerged as a new state-of-the-art deep generative model \cite{diffsurvey}. The fundamental concept of the diffusion model entails systematically perturbing the distribution of data during the forward diffusion process by introducing Gaussian noise. Subsequently, the data distribution is recovered through the reverse diffusion process, which can be viewed as a denoising procedure. Specifically, within the forward diffusion process, by iteratively adding Gaussian noise $T$ times to any initial sample ${{\mathbf{x}} _0}$, we can obtain ${{\mathbf{x}} _1},{{\mathbf{x}} _2},...,{{\mathbf{x}} _T}$. As $T$ approaches infinity, the original features of sample ${{\mathbf{x}}  _0}$ completely vanish and become pure Gaussian noise. This process can be represented as follows:
         \begin{equation}\label{}
         q\left( {{{\mathbf{x}}_1},...,{{\bf{x}}_T}\mid {{\bf{x}}_0}} \right) = \mathop \prod \limits_{t = 1}^T q\left( {{{\bf{x}}_t}\mid {{\bf{x}}_{t - 1}}} \right),
         \end{equation}
         \begin{equation}\label{dif1}
         q\left( {{{\bf{x}}_t}\mid {{\bf{x}}_{t - 1}}} \right): = {\cal N}\left( {{{\bf{x}}_t};\sqrt {1 - {\beta _t}} {{\bf{x}}_{t - 1}},{\beta _t}{\bf{I}}} \right),
         \end{equation}        
    where, $\beta _t$ is a parameter that controls the progress of noise. From \eqref{dif1}, it can be inferred that given the sample ${\bf{x}}_{t - 1}$, the sample ${\bf{x}}_t$ at time $t$ follows a Gaussian distribution with a mean of $\sqrt {1 - {\beta _t}} {{\bf{x}}_{t - 1}}$ and a variance of ${\beta _t}{\bf{I}}$. The parameters under this condition only depend on the ${\bf{x}}_{t{\rm{ - }}1}$ at the previous time step. Therefore, the diffusion process is a Markov process.
    
    When $\beta _t$ is sufficiently small, the reverse diffusion process $q\left( {{{\bf{x}}_{t{\rm{ - }}1}}\mid {{\bf{x}}_t},{{\bf{x}}_0}} \right)$ is the posterior probability distribution of the forward diffusion process $q\left( {{{\bf{x}}_t}\mid {{\bf{x}}_{t - 1}}} \right)$. In order to achieve incremental sampling from Gaussian noise ${{\bf{x}}_T}$ to obtain real samples, it is necessary for the generative model ${p_\theta }\left( {{{\bf{x}}_{0:T}}} \right)$ to learn sufficiently good parameters $\theta $ from the training samples. This process can be represented as follows:
         \begin{equation}\label{}
         {p_\theta }\left( {{{\bf{x}}_{0:T}}} \right) = p\left( {{{\bf{x}}_T}} \right)\prod\limits_{t = 1}^T {{p_\theta }} \left( {{{\bf{x}}_{t - 1}}\mid {{\bf{x}}_t}} \right),
         \end{equation}
         \begin{equation}\label{}
         {p_\theta }\left( {{{\bf{x}}_{t - 1}}\mid {{\bf{x}}_t}} \right) = {\cal N}\left( {{{\bf{x}}_{t - 1}};{\mu _\theta }\left( {{{\bf{x}}_t},t} \right),{\Sigma _\theta }\left( {{{\bf{x}}_t},t} \right)} \right),
         \end{equation}
    where, $p\left( {{{\bf{x}}_T}} \right){\rm{ = }}{\cal N}\left( {{{\bf{x}}_T};0,{\bf{I}}} \right)$. Finally, the reverse diffusion process can be achieved by utilizing a well-trained ${p_\theta }\left( {{{\bf{x}}_{t - 1}}\mid {{\bf{x}}_t}} \right)$ to approximate $q\left( {{{\bf{x}}_{t{\rm{ - }}1}}\mid {{\bf{x}}_t},{{\bf{x}}_0}} \right)$.
    
\begin{figure*}[ht]
	\centering
	\includegraphics[width=0.9\textwidth]{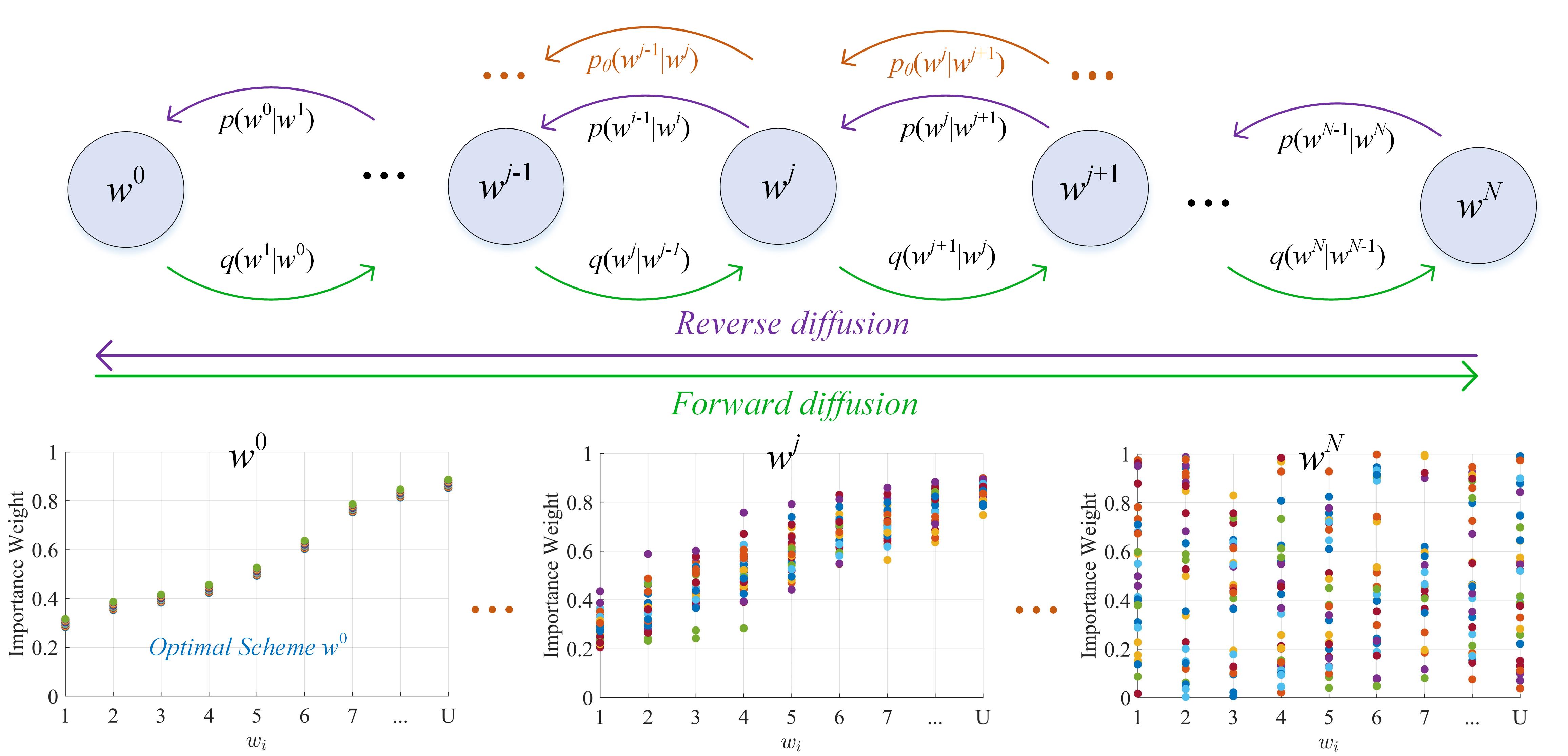}%
	\caption{The design principles of diffusion model. The diffusion model performs multi-step denoising on noise and generates an optimal allocation scheme.}
	\label{diffusionpr}
	\end{figure*}

    \textbf{AI-Generated Power Allocation Scheme: }Motivated by diffusion model-based AI-generated contract \cite{Du2023AIGeneratedIM}, we propose an AI-generated power allocation scheme. In Deep Reinforcement Learning (DRL), the intelligent agent learns the optimal policy through interaction with the environment to maximize cumulative rewards. As shown in Fig. \ref{diffusionpr}, the AI-generated algorithm is capable of addressing the challenges posed by high-dimensional state spaces and complex action spaces. Compared to the aforementioned two methods, i.e., Avg-SemCom and Conf-SemCom, this method exhibits superior performance in generating resource allocation schemes. Specifically, we represent the environment using the vector $\bm{e}$, which encompasses various factors such as the wireless channel model, the transmission power $P$, and the number of objects $U$ involved in the semantic communication. In this given environment, our objective is to maximize the expected cumulative reward across a series of time steps, aiming to determine the transmission power weights ${\bm{w}_i}$ for each object.

    We first construct a generative model ${\pi _\theta }\left( {\bm{w}\mid \bm{e}} \right)$ that can map the environmental state $\bm{e}$. The reverse process of the conditional diffusion model can be represented as follows:
         \begin{equation}\label{}
         \begin{split}  
         {\pi _\theta }\left( {\bm{w}\mid \bm{e}} \right) & = {p_\theta }\left( {{\bm{w}^{0:N}}\mid \bm{e}} \right) \\ &= {\cal N}\left( {{\bm{w}^N};0,{\bf{I}}} \right)\mathop \prod \limits_{i = 1}^N {p_\theta }\left( {{\bm{w}^{j - 1}}\mid {\bm{w}^j}\bm{,e}} \right),
         \end{split}
         \end{equation}   
    where, ${p_\theta }\left( {{\bm{w}^{j - 1}}\mid {\bm{w}^j}\bm{,e}} \right)$ can be modeled as a Gaussian distribution ${\cal N}\left( {{\bm{w}^{_{j - 1}}};{\mu _\theta }\left( {{\bm{w}^j},\bm{e},j} \right),{\Sigma _\theta }\left( {{\bm{w}^j},\bm{e},j} \right)} \right)$. According to Denoising Diffusion Probabilistic Models (DDPM) \cite{ddpm2020}, the covariance matrix ${\Sigma _\theta }\left( {{\bm{w}^j},\bm{e},j} \right)$ of this Gaussian distribution is ${\beta _j}{\bf{I}}$, and the mean ${\bm{\mu }_\theta }\left( {{\bm{w}^j},\bm{e},j} \right)$ can be represented as $\frac{1}{{\sqrt {{\alpha _j}} }}\left( {{\bm{w}^j} - \frac{{{\beta _j}}}{{\sqrt {1 - {{\bar \alpha }_j}} }}{\varepsilon _\theta }\left( {{\bm{w}^j},\bm{e},j} \right)} \right)$. Initially, we sample ${\bm{w}^N} \sim {\cal N}\left( {0,{\bf{I}}} \right)$, and then from the reverse diffusion chain parameterized by $\theta $ as:
         \begin{equation}\label{dif2}
         {\bm{w}^{j - 1}}|{\bm{w}^j} = \frac{1}{{\sqrt {{\alpha _j}} }}\left( {{\bm{w}^j} - \frac{{{\beta _j}}}{{\sqrt {1 - {{\bar \alpha }_j}} }}{\varepsilon _\theta }\left( {{\bm{w}^j},\bm{e},j} \right)} \right) + \sqrt {{\beta _j}} \varepsilon .
         \end{equation}
    From \eqref{dif2}, as can be seen that the result is only related to ${\bm{w}^j}$ and the added noise $\varepsilon $. Therefore, the training of the denoising process ${\pi _\theta }$ can be achieved by training ${\varepsilon _\theta }$. Subsequently, we use the quality network ${Q_v}$ to train the ${\varepsilon _\theta }$, which represents the expected cumulative reward that an agent takes an allocation scheme in the current state and executes accordingly. The objective function that needs to be optimized becomes:
        \begin{equation}\label{}
        \pi=\underset{\pi_{\theta}}{\arg \min } \mathcal{L}(\theta)=-\mathbb{E}_{\boldsymbol{w}^{0} \sim \pi_{\theta}}\left[Q_{v}\left(\boldsymbol{e}, \boldsymbol{w}^{0}\right)\right].
        \end{equation}
    Utilizing the double Q-learning method \cite{Hasselt2010DoubleQ}, the network ${Q_v}$ is learned by minimizing the Bellman operator. Consequently, we construct two networks, specifically ${Q_{v_1}}$ and ${Q_{v_2}}$, along with corresponding target networks ${Q_{v_1^{'}}}$, ${Q_{v_2^{'}}}$, and ${\pi _{{\theta ^{'}}}}$. The optimization of ${v_1}$ and ${v_2}$ is achieved by minimizing the objective
        \begin{equation}\label{}
        \mathbb{E}_{\boldsymbol{w}_{t+1}^{0} \sim \pi_{{\theta}^{\prime}}}
        \left[ {{{\left\| \begin{array}{l}
        \left( {r\left( {\bm{e},{\bm{w}_t}} \right) + \gamma \mathop {\min }\limits_{i = 1,2} {Q_{v_i^\prime }}\left( {\bm{e},\bm{w}_{t + 1}^0} \right)} \right)\\
        - {Q_{{v_i}}}\left( {\bm{e},{\bm{w}_t}} \right)
        \end{array} \right\|}^2}} \right].    
        \end{equation}
    In DRL, we set the training parameters: batch size ${N_b}$, discount factor $\gamma $, diffusion step $N$, soft target update parameter $\tau $, and exploration noise $\epsilon $. The loss function can be represented as
        \begin{equation}\label{loss}
        \mathcal{L} \text{= }\frac{1}{{{N}_{b}}}{{\sum\nolimits_{j}{\left( {{r}_{j}}+\gamma {{Q}^{'}}_{{{v}^{'}}}\left( {{\bm{e}}_{j}},{{\bm{w}}^{'}}_{t}^{o} \right)-{{Q}_{v}}\left( {{\bm{e}}_{j}},{{\bm{w}}_{j}} \right) \right)}}^{2}}.
        \end{equation}
    Then, we can obtain the optimal allocation scheme based on the wireless communication environment. The detail of the AI-generated scheme is shown in Algorithm \ref{diffalgo}. 
    	\begin{algorithm}[ht]
		\caption{Diffusion Model-based AI-Generated Scheme.} 
		\label{diffalgo}
		\hspace*{0.02in} {\bf Training:}         
		\begin{algorithmic}[1] 
                \State {\bf Initial:}
			\State \quad Initialize replay buffer R and the weights of models, i.e., $\theta$, ${\theta ^{'}}$, $v$, ${v^{'}}$
			\For{{\rm{Episode  =  1}} to {Max\_episode}} 
			\State Initialize a random process N
                \For{{\rm{Step  =  1}} to {Max\_step}}
                \State Observe the existing environment ${\bm{e}_t}$
                \State According to \eqref{dif2}, set $\bm{w}_t^{N}$ as Gaussian noise and generate allocate scheme $\bm{w}_t^0$ by denoising $\bm{w}_t^N$ using ${\varepsilon _\theta }$ 
                \State Combine $\bm{w}_t^0$ with the exploration noise $\epsilon $ 
                \State Execute scheme $\bm{w}_t^0$ and observe reward score according to \eqref{reward}
                \State Save the record (${{\bm{e}}_{t}}$, $\bm{w}_t^{0}$, ${{\tau }_{t}}$) in R
                \State Randomly sample ${{N}_{b}}$ records (${{\bm{e}}_{j}}$, ${\bm{w}_{j}}$, ${{\tau }_{j}}$) from R as a minibatch
                \State Minimize the loss to update ${{Q}_{v}}$ according to \eqref{loss}
                \State Update ${{\varepsilon }_{\theta }}$ by taking gradient descent step on
                \State \quad${{\nabla }_{\theta }}{{\varepsilon }_{\theta }}\approx \frac{1}{{{N}_{b}}}\sum\nolimits_{j}{{{\nabla }_{{\bm{w}^{0}}}}}{{Q}_{v}}\left( \bm{e,}{{\bm{w}}^{0}} \right){{|}_{\bm{e=}{{\bm{e}}_{j}}}}{{\nabla }_{\theta }}{{\varepsilon }_{\theta }}|{{\bm{e}}_{j}}$               
                \State ${{\theta }^{\prime }}\leftarrow \tau \theta +(1-\tau ){{\theta }^{\prime }}$
                \State ${{v}^{\prime }}\leftarrow \tau v+(1-\tau ){{v}^{\prime }}$                 
                \EndFor
			\EndFor 
                \State\Return{${{\varepsilon }_{\theta }}$}              
		\end{algorithmic}
            \hspace*{0.02in} {\bf Inference:}              
		\begin{algorithmic}[1]
                \State Input $\bm{e}$
                \State According to \eqref{dif2}, denoise Gaussian noise using ${{\varepsilon }_{\theta }}$ to generate the optimal allocation scheme $\bm{w}_{{}}^{0}$
                \State\Return {The optimal resource allocate scheme $\bm{w}_{{}}^{0}$}
            \end{algorithmic}
	\end{algorithm}
    
	\section{Numerical Results}\label{S5}
    This section primarily presents the experimental settings, materials, and results in this study. Initially, we evaluate the effectiveness of the improvements made to the object detector YOLOv7-X and compare its performance with other state-of-the-art models. Subsequently, we assess the performance of the proposed YOLO-based semantic communication system, verify the cost savings of semantic communication, and examine the impact of two proposed power allocation schemes on the transmission quality for critical information.
        \subsection{Environment Setup}
\begin{figure}[ht]
	\centering
	\includegraphics[width=0.4\textwidth]{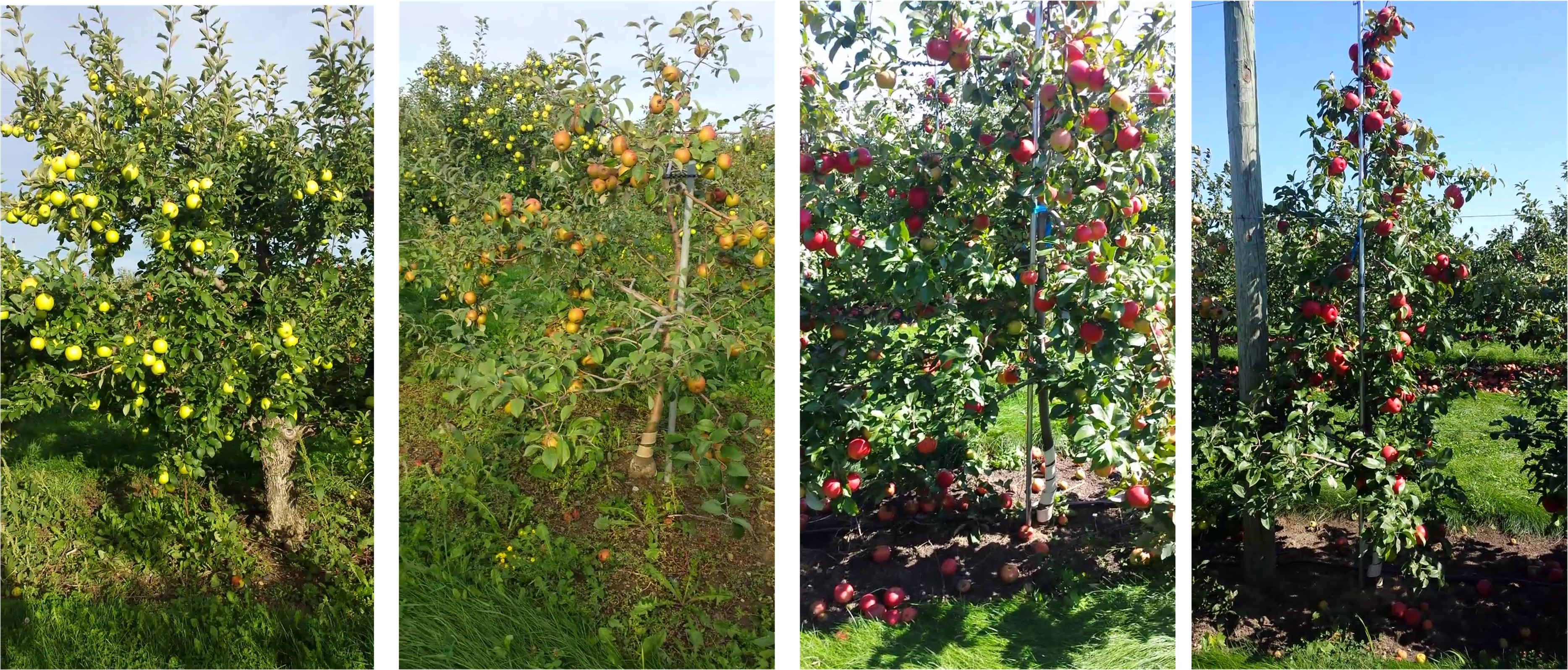}%
	\caption{A partial overview of MinneApple dataset.}
	\label{minne}
	\end{figure}

    The experimental platform is built on a generic Ubuntu 20.04 system with 2 Intel(R) Xeon(R) Silver 4110 CPUs and GeForce RTX 3090 GPU. The parameters of the object detection model training process used are shown in TABLE \ref{trainpas}. The MinneApple dataset~\cite{41Hni2019MinneAppleAB} is the apple image dataset used in this experiment. It is a publicly available dataset utilized  for apple detection and segmentation, containing images of multiple apple varieties at different stages of growth, with a large number of densely packed small apples. The MinneApple dataset contains a total of 670 labeled images and 331 unlabeled images. Fig. \ref{minne} shows example images from the MinneApple dataset.
        \begin{table}[!ht]
        \centering
        \caption{Object Detection Models Training Parameters}
        \begin{tabular}{cccccccccc}
        \hline
            Parameter & Value & Parameter & Value \\ \hline
            Learning Rate & 0.01 & Epochs & 300 \\ 
            Batch Size & 16 & Momentum & 0.937  \\ 
            Image Size & 640$ \times $640  & Weight Decay & 0.0005  \\ \hline
        \end{tabular}
        \label{trainpas}
        \end{table}

    We use the Fisher-Snedecor ${\cal F}$ channel model~\cite{33Kang2022PersonalizedSI} in wireless semantic communication to analyze the performance of our model. The small-scale fading between the UAV and users is represented by the Fisher-Snedecor ${\cal F}$ fading distribution, while small-scale variations follow the Nakagami$ - m$ distribution and shadowing follows the inverse Nakagami$ - m$ distribution \cite{33Kang2022PersonalizedSI}. We set the fading parameter ${m_f}$ = 6, the shadowing parameter ${m_s}$ = 6 and the transmit power $P$ = 3000 W by default. In addition, the parameters of the resource allocation scheme generated by the AI-generated algorithm during the training process are shown in TABLE \ref{difftrain}.

    \begin{table}[!ht]
    \centering
    \caption{Diffusion Model-Based AI-Generated Algorithm \\Training Parameters}
    \begin{tabular}{cc}
    \hline
        Parameter & Value  \\ \hline
        Diffusion Step $N$ &  50 \\ 
        Batch Size $N_b$ & 512  \\ 
        Discount Factor $\gamma $ & 0.95  \\ 
        Soft Target Update Parameter $\tau $ & 0.005  \\ 
        Exploration Noise $\epsilon $ & 0.05  \\ 
        The Learning Rate of Network ${\varepsilon _\theta }$ & ${10^{-5}}$  \\ 
        The Learning Rate of Network $Q_{v}$ & ${10^{-4}}$  \\ \hline
    \end{tabular}
    \label{difftrain}
    \end{table}

        \subsection{Results and Analysis}
    \textbf{Results of ablation experiment. }To evaluate the effectiveness of the ELAN-H  and SimAM attention modules, we utilize the amount of parameters and computational complexity, i.e., Floating Point Operations (FLOPs), as well as AP@0.5 and AP@0.5:0.95 as indicators to measure the performance of the models. Where AP@0.5 and AP@0.5:0.95 are commonly-used evaluation standards in object detection, with higher values indicating better model performance. From the results in Table \ref{meds}, the utilization of the ELAN-H module leads to 1.3\% and 1.7\% increases in AP@0.5 and AP@0.5:0.95, respectively, while reducing the amount of parameters by 24\% and FLOPs by 19\%. The incorporation of the SimAM attention module enhances the value of AP@0.5 by 0.8\%, with no changes to the amount of parameters and FLOPs.
        \begin{table*}[htbp]
            \centering
            \caption{Results of Ablation Experiment.}
            \begin{tabular}{ccccc}
            \hline
                Model & Parameters & FLOPs & AP@0.5 & AP@0.5:0.95  \\ \hline
                YOLOv7-X & 70.7M & 188.0G & 87.8\% & 43.7\%  \\ 
                YOLOv7-X+ELAN-H & 53.5M & 152.6G & 89.1\% & 45.4\%  \\ 
                YOLOv7-X+ELAN-H+SimAM & 53.5M & 152.6G & \textbf{89.8\%} & \textbf{45.4\%}  \\ \hline
            \end{tabular}
            \label{meds}
        \end{table*}
        
    \textbf{Comparison with other state-of-the-art object detectors. }We conduct a performance comparison of our enhanced YOLOv7-X model with other advanced object detection models. In addition to the comparison items presented in Table \ref{meds}, we also compare the detection speed.
        \begin{table*}[htbp]
            \centering
            \caption{Comparison of State-of-the-art Object Detectors.}
            \begin{tabular}{cccccccccc}
            \hline
                Model & Parameters & FLOPs & AP@0.5 & AP@0.5:0.95 & FPS  \\ \hline
                Faster R-CNN~\cite{18Ren2015FasterRT}  & 41.7M & 59.4G & 74.1\% & 33.1\% & 34 \\ 
                RetinaNet~\cite{42Lin2017FocalLF}  & 56.9M & 74.4G & 62.4\% & 24.8\% & 32 \\ 
                FCOS~\cite{43Tian2019FCOSFC}  & 51.2M & 66.3G & 66.6\% & 27.8\% & 25  \\ 
                Scaled-YOLOv4-p5~\cite{44Wang2020ScaledYOLOv4SC}  & 70.2M & 165.1G & 88.3\% & 45.9\% & 22   \\ 
                YOLOX-X~\cite{34Ge2021YOLOXEY}  & 104.5M & 312.0G & 88.4\% & \textbf{47.0\%} & 27 \\ 
                YOLOv5-X & 86.2M & 203.8G & 87.5\% & 44.7\% & 32  \\ 
                YOLOR-CSP-X~\cite{45Wang2021YouOL}]  & 96.4M & 225.5G & 81.3\% & 42.0\% & 22   \\ 
                PPYOLOE-X~\cite{46Xu2022PPYOLOEAE}  & 95.3M & 204.9G & 88.6\% & 45.4\% & 19   \\ 
                Ours & 53.5M & 152.6G & \textbf{89.8\%} & 45.4\% & \textbf{34} \\ \hline
            \end{tabular}
            \label{compare}
        \end{table*}       
    As shown in Table \ref{compare}, our proposed model achieves the best performance in terms of AP@0.5 compared to other models. Although the performance difference between our proposed model and other advanced YOLO series models is not significant, our model greatly reduces the amount of parameters and FLOPs while achieving the fastest detection speed. However, Faster R-CNN, RetinaNet, and FCOS models, although having smaller parameter size, perform poorly on the MinneApple dataset, which contains many small objects, and their detection performance fails to meet the requirements of our proposed scenarios.
    
\begin{figure}[ht]
	\centering
        \includegraphics[width=0.45\textwidth]{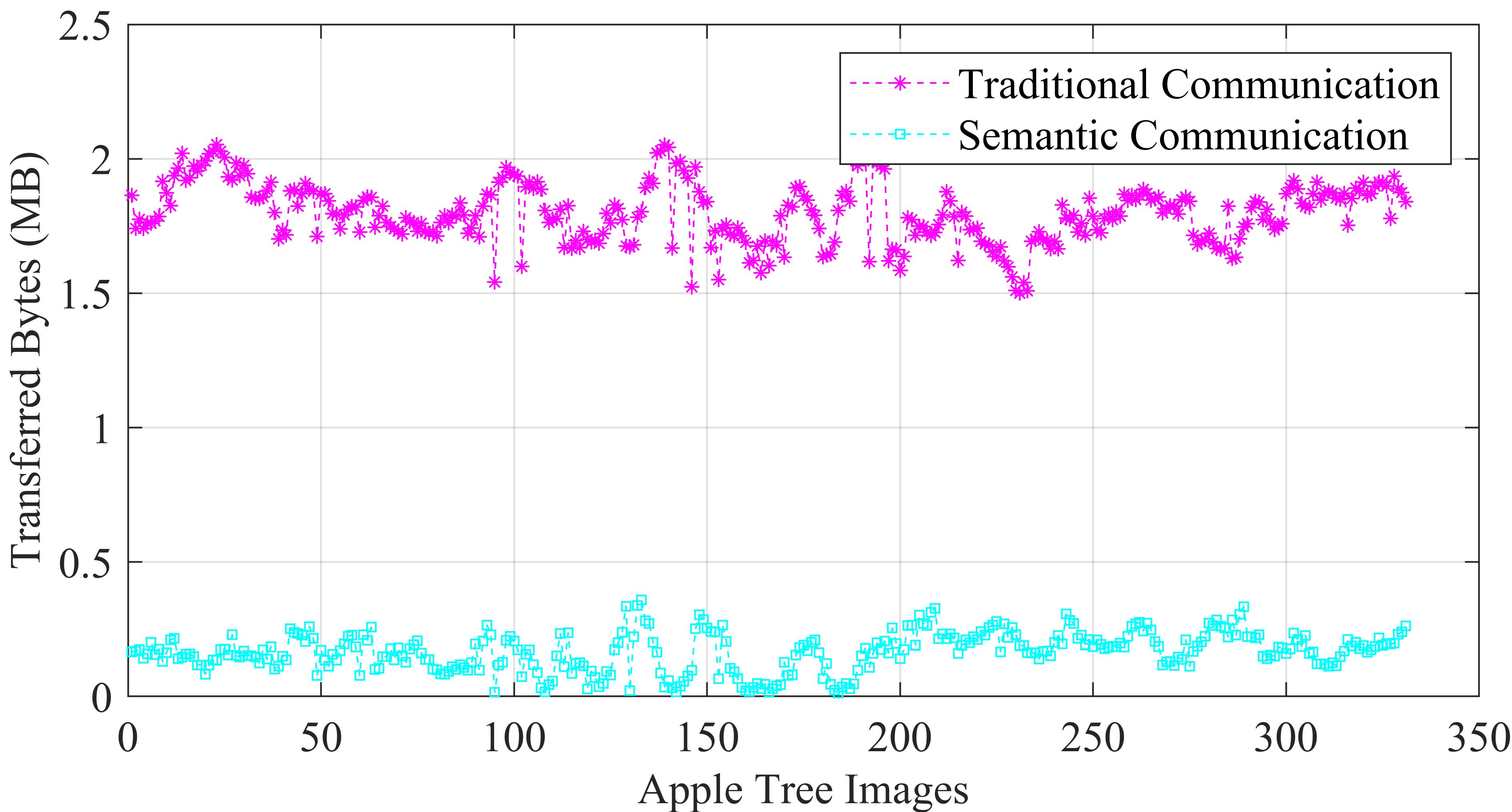}%
	\caption{Comparison of transferred bytes for each image in the two communication methods.}
	\label{reduce}
	\end{figure}

    \textbf{The effects of reducing communication overhead. }We use 331 unannotated test images from the MinneApple dataset as the images to be sent by the UAV. Fig. \ref{reduce} illustrates a comparison between the data size of images transmitted through conventional communication methods and the data size resulting from semantic communication after the implementation of semantic feature extraction. The aggregate size of the original images amounts to 595.2MB. However, following semantic feature extraction, the volume of data required for transmission by edge devices is considerably reduced to 55.4MB, encompassing 54.8MB of image format data and 0.6MB of text format data. This reduction corresponds to a 91\% decrease in communication costs, thereby substantially minimizing power consumption during transmission.

\begin{figure}[htbp]
        \centering 
        \subfigure[Transmission Distance $D = 10m.$]{
        \label{10m}
        \includegraphics[width=0.45\textwidth,height = 4cm]{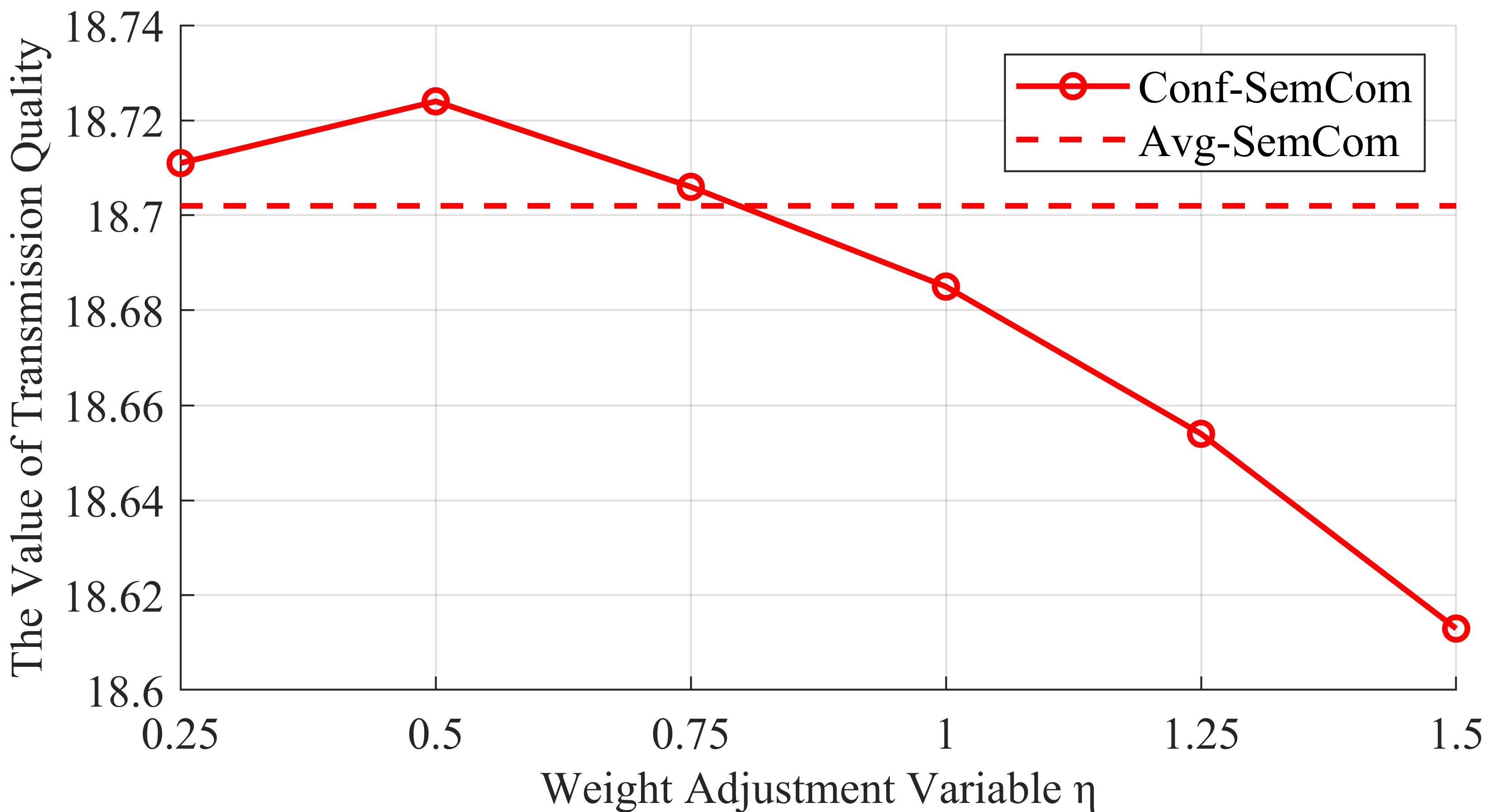}}
        \subfigure[Transmission Distance $D = 20m.$]{
        \label{20m}
        \includegraphics[width=0.45\textwidth,height = 4cm]{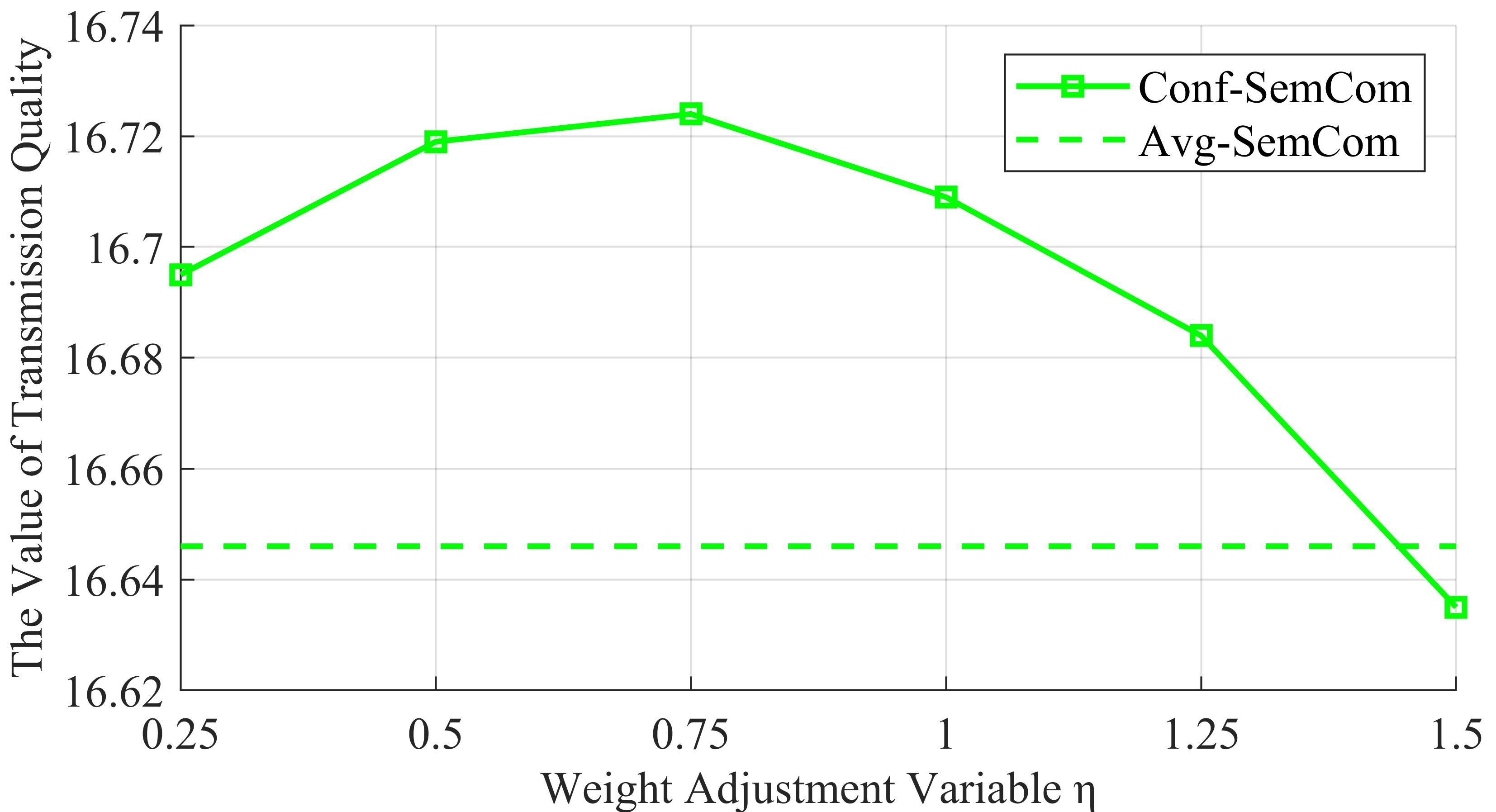}}
        \subfigure[Transmission Distance $D = 30m.$]{
        \label{30m}
        \includegraphics[width=0.45\textwidth,height = 4cm]{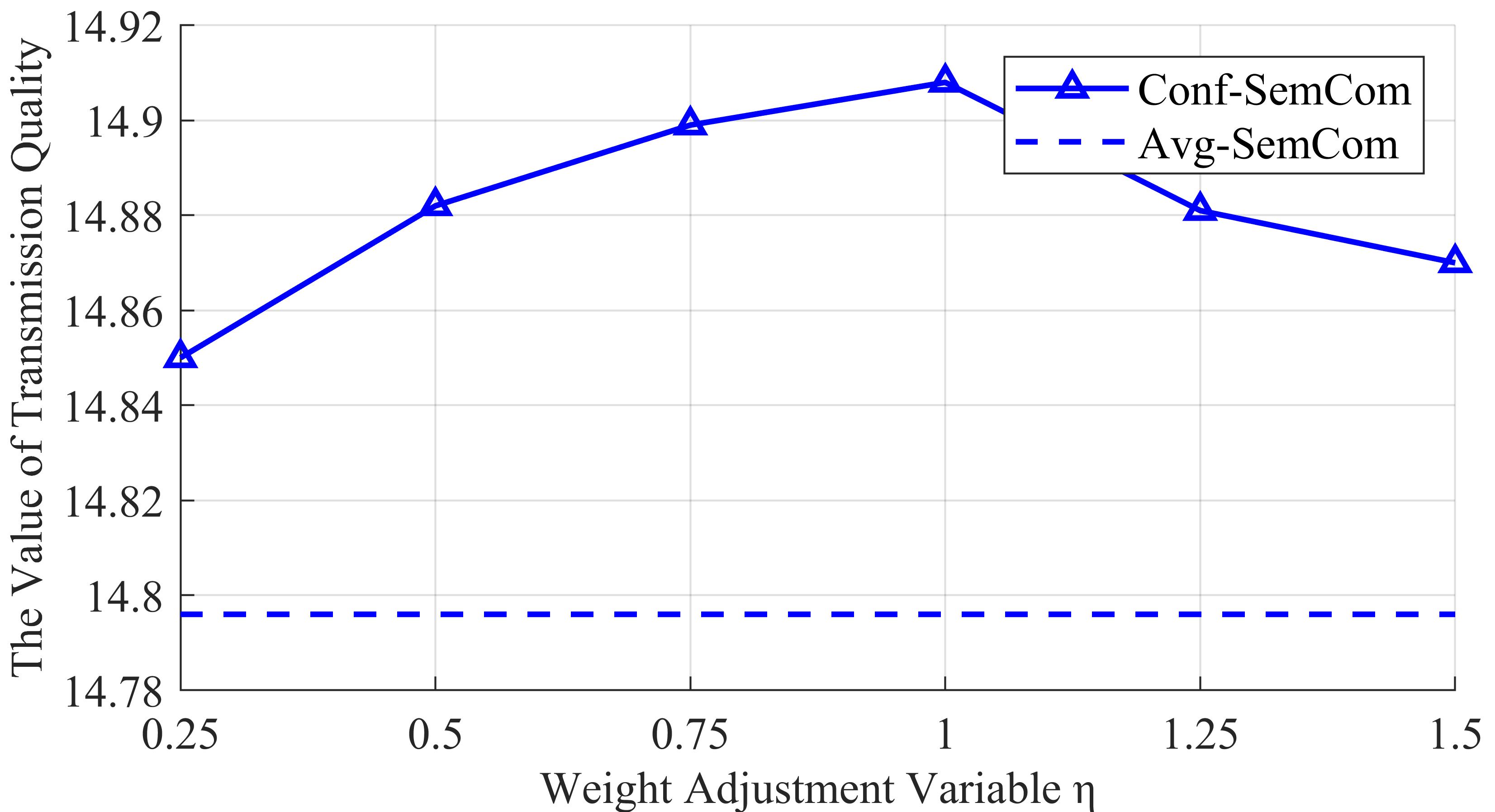}}
        \caption{The curves of transmission quality scores with different weight adjustment variables $\eta $ and transmission distance $D$.}
        \label{var}
        \end{figure}

    \textbf{The effect of $\eta $ on Conf-SemCom. }Given that the textual data required for transmission by the UAV is considerably smaller than image data, we focus solely on the impact of wireless transmission environments on the semantic information transfer of image formats. Taking Fig. \ref{apple351} as an example, the UAV first detects that the image contains 30 objects and then allocates power for transmission according to the confidence of each object. We first show the effectiveness of the proposed Conf-SemCom method. Additionally, to investigate the impact of variable $\eta $ on the transmission performance, we increase $\eta $ from 0.25 to 1.5 and the transmission distance from 10m to 30m. Each experiment is repeated 100 times, and the average results are shown. The curves of transmission quality values (i.e., MIST scores) is shown in Fig. \ref{var}. 
    It can be observed that due to allocating more resources to important information, Conf-SmeCom outperforms the Avg-SemCom method in most cases, with its effectiveness becoming more evident as the transmission distance increases. Furthermore, the optimal value of variable $\eta $ varies across different transmission distances. At transmission distances of 10m, 20m, and 30m, the highest MIST scores are achieved when $\eta $ equals 0.5, 0.75, and 1, respectively. This indicates that appropriately increasing the value of variable $\eta $ as the transmission distance increases, while keeping the total power $P$ unchanged, allows for a better enhancement of overall communication quality by increasing the transmission power $p_i$ for more significant objects $i$.
    
\begin{figure}[ht]
	\centering
	\includegraphics[width=0.45\textwidth]{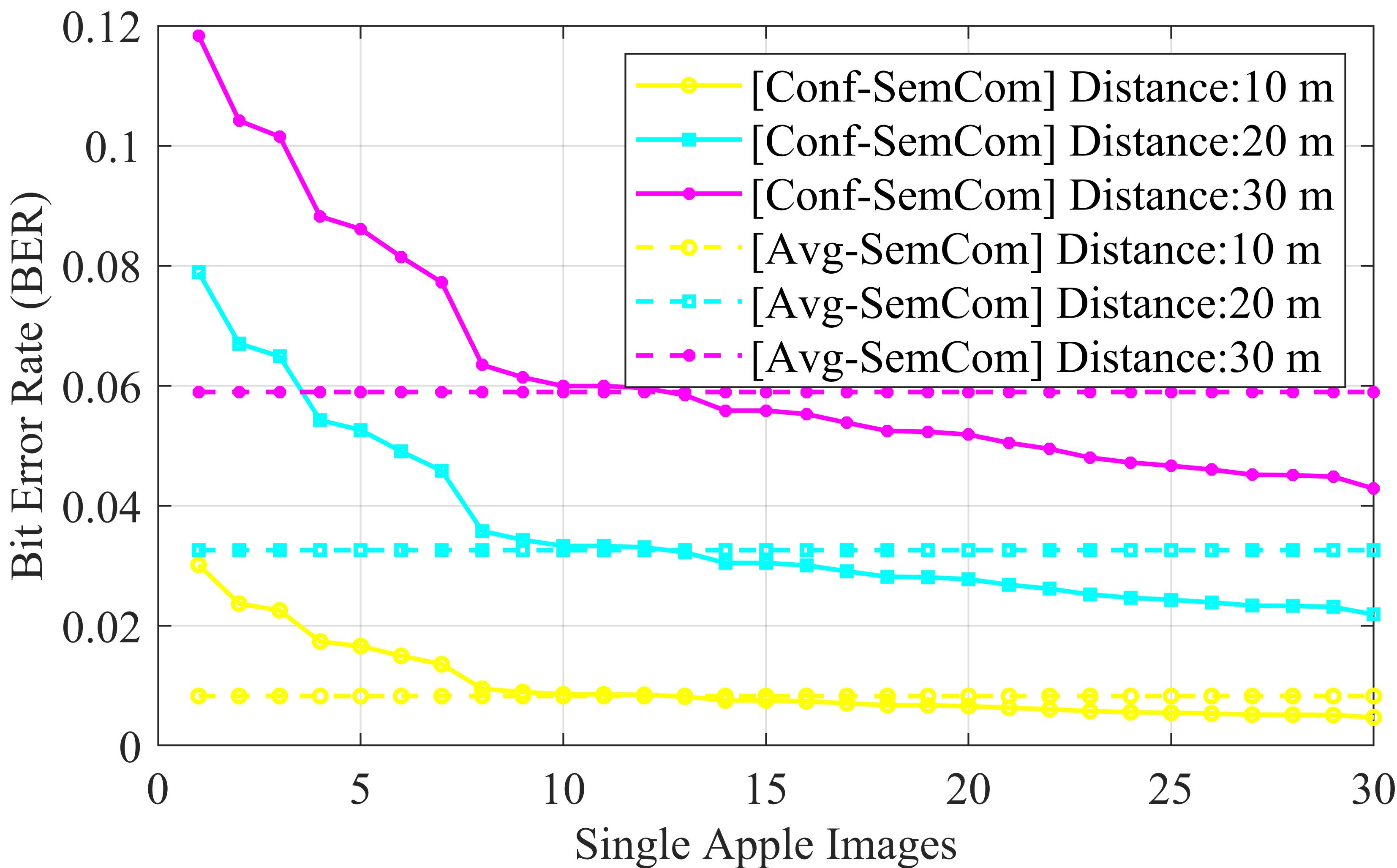}%
	\caption{Bit error rate of images at different transmission distances under two transmission methods.}
	\label{allo}
	\end{figure}

    \textbf{The effects of power allocation. }We investigate the impact of two power allocation methods, i.e., Avg-SemCom and Conf-SemCom, on the transmission quality of images with different levels of confidence, over different transmission distances (i.e., 10m, 20m, and 30m). As shown in Fig. \ref{allo}, the horizontal axis represents the 30 detected images, sorted in ascending order of confidence levels, and the vertical axis represents the Bit Error Rate (BER) values derived from the channel model according to the allocated power. It is evident that, as the distance increases, the image transmission quality declines for both communication methods. However, Conf-SemCom opts to allocate more power to semantically important information, resulting in reduced error rates for crucial semantic information even under poor channel conditions.
 
\begin{figure*}[!htbp]
	\centering
	\includegraphics[width=0.9\textwidth]{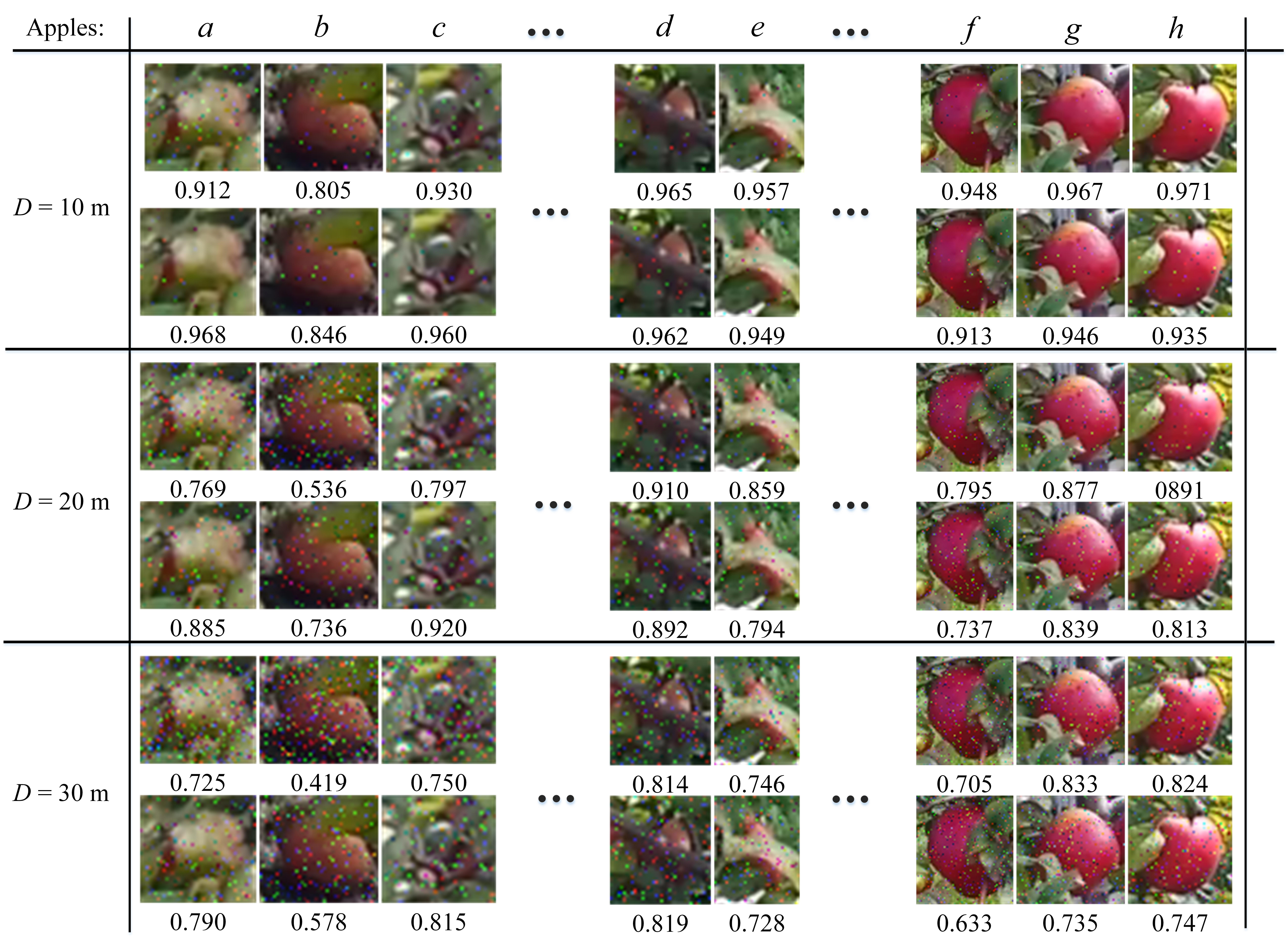}%
	\caption{The transmission effects and corresponding SSIM values of partial images at different transmission distances. The odd-numbered rows are Conf-SemCom and the even-numbered rows are Avg-SemCom.}
	\label{ssim}
	\end{figure*}

    To facilitate a comprehensive comparison between the two power allocation methods, Fig. \ref{ssim} illustrates the transmission performance of select images at various communication distances. We use SSIM to evaluate the transmission quality of each object. From the Fig. \ref{ssim}, the transmitted image quality for both methods degrades significantly as the transmission distance increases. Moreover, due to the uniformed distribution of transmission power, the SSIM values transmitted by Avg-SemCom for each image exhibit a relatively uniform and irregular pattern, resulting in certain critical images possessing inferior transmission quality compared to original images. For instance, apples $a$ and $c$ are heavily occluded and contain very little usable information. The significance of semantic features for these images is less than that for other images. However, their transmission quality surpasses that of other images, which is unreasonable. In contrast, Conf-SemCom allocates increased power to salient images, enabling the high-quality transmission of these images even in poor channel conditions. 

\begin{figure}[!ht]
	\centering
	\includegraphics[width=0.45\textwidth]{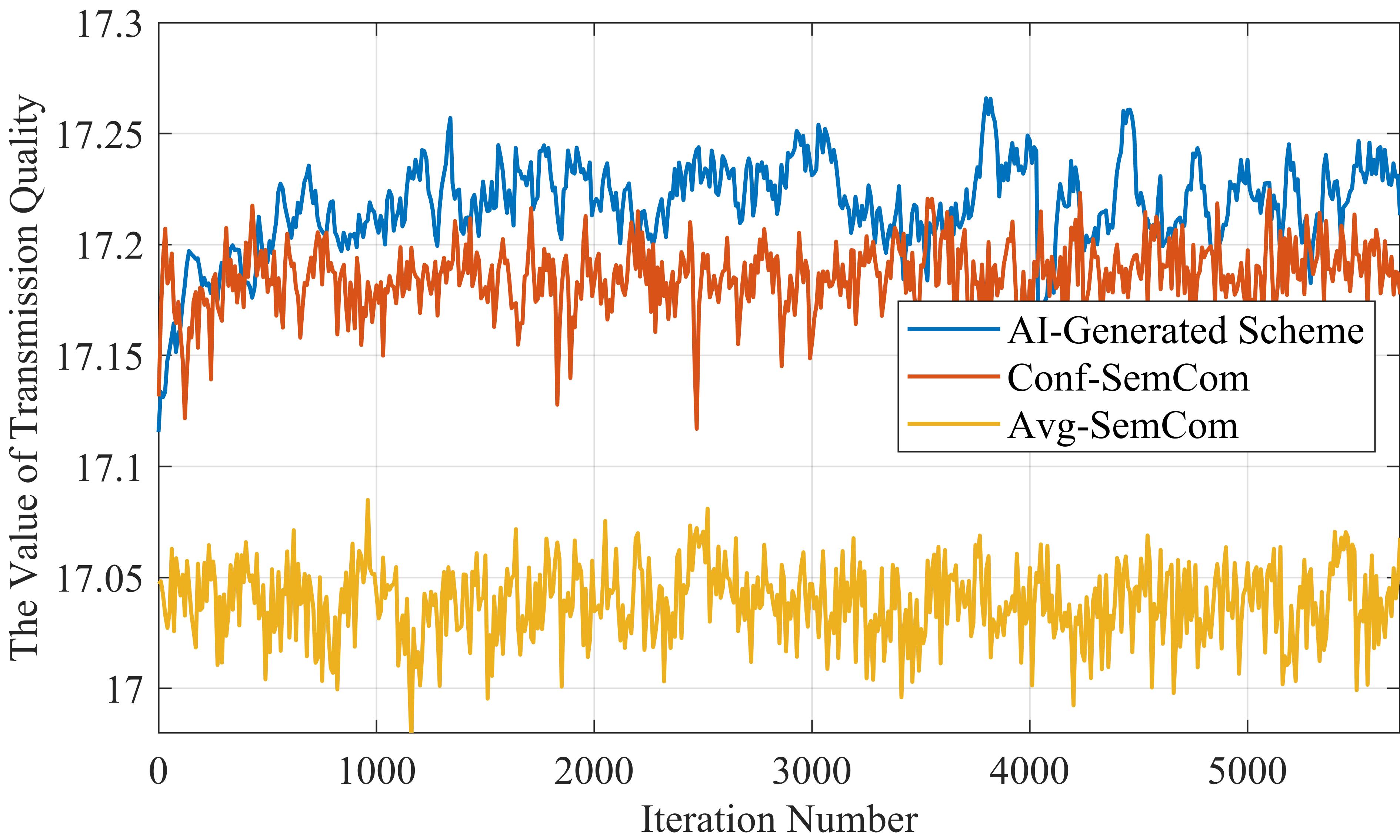}%
	\caption{Comparison of the training process of the diffusion model-based AI-generated algorithm and the results of two other methods (i.e., Avg-SemCom and Conf-SemCom), with transmission distance $D$ = 20 m and transmission power $P$ = 4 kW.}
	\label{aischeme}
	\end{figure}

    \textbf{The effects of AI-Generated scheme. }We compare the diffusion model-based AI-generated scheme with two other transmission power allocation methods for transmitting semantic information at the transmission distance $D$ = 20 m and the transmission power $P$ = 4 kW. As illustrated in Fig. \ref{aischeme}, the diffusion model-based AI-generated algorithm exhibits rapid training speed during the optimization of power allocation schemes, surpassing the confidence-based allocation scheme at approximately 500 iterations. The superiority of the AI-generated approach primarily stems from the exploration conducted through diffusion method, which enhances the flexibility of strategies and prevents the model from getting trapped in suboptimal solutions. Furthermore, it is evident that the Avg-SemCom method significantly underperforms in terms of the MIST score compared to the other two schemes, indicating the necessity of considering the importance of semantic information during the communication process. 
        \section{Conclusion}\label{S6}
    In this paper, we have proposed a YOLO-based semantic communication framework for developing a virtual apple orchard case, focusing on optimizing semantic information transmission and resource allocation for images collected by edge devices. Initially, we have enhanced the performance of the object detector YOLOv7-X on a real apple dataset and have employed the optimized object detector to extract semantic information from images captured by edge devices, aiming to reduce transmission costs. Furthermore, to ensure the high-quality transmission of essential semantic information, we have allocated resource based on the significance of their semantic content. Specifically, we have allocated the transmission power of semantic information based on the confidence generated by the object detection algorithm and the scheme generated by the diffusion model-based AI-generated algorithm, respectively. Numerical results have demonstrated that the proposed framework and strategy have considerably reduced communication costs and have markedly improved the transmission quality of important information during communication.
\bibliographystyle{IEEEtran}
\bibliography{IEEEabrv,Ref}
\end{document}